\documentclass[pre,aps,reprint,amsmath,amssymb,superscriptaddress]{revtex4-1}
\usepackage[utf8]{inputenc}
\usepackage{graphicx,amsmath,amssymb}
\usepackage[dvipsnames]{xcolor}
\usepackage[colorlinks=true,citecolor=ForestGreen,linkcolor=Red,urlcolor=Blue,hypertexnames=false]{hyperref}
\newcommand{\ie}{\emph{i.e.}}
\newcommand{\eg}{\emph{e.g.}}
\newcommand{\ee}{\text{e}}
\newcommand{\avg}[1]{\langle #1\rangle}
\newcommand{\abs}[1]{\vert #1\vert}
\newcommand{\GT}{\mathcal{G}}
\newcommand{\COM}{\mathsf{C}}
\newcommand{\TRU}{\mathsf{T}}
\newcommand{\CF}{\mathcal{C}}
\newcommand{\neff}{n_{\rm eff}}
\newcommand{\kavg}{\overline{k}}
\newcommand{\figSI}[1]{\includegraphics[scale = 0.7]{#1}}
\begin{document}
\title{Optimal timescale for community detection in growing networks}
\author{Matúš Medo}
\email{matus.medo@unifr.ch}
\affiliation{Institute of Fundamental and Frontier Sciences, University of Electronic Science and Technology of China, Chengdu 610054, PR China}
\affiliation{Department of Physics, University of Fribourg, 1700 Fribourg, Switzerland}
\author{An Zeng}
\affiliation{School of Systems Science, Beijing Normal University, Beijing, 100875, P. R. China}
\author{Yi-Cheng Zhang}
\affiliation{Department of Physics, University of Fribourg, 1700 Fribourg, Switzerland}
\author{Manuel S. Mariani}
\affiliation{Institute of Fundamental and Frontier Sciences, University of Electronic Science and Technology of China, Chengdu 610054, PR China}
\affiliation{URPP Social Networks, Universit\"at Z\"urich, Switzerland}
\affiliation{Department of Physics, University of Fribourg, 1700 Fribourg, Switzerland}
\date{\today}

\begin{abstract}
Time-stamped data are increasingly available for many social, economic, and information systems that can be represented as networks growing with time. The World Wide Web, social contact networks, and citation networks of scientific papers and online news articles, for example, are of this kind. Static methods can be inadequate for the analysis of growing networks as they miss essential information on the system's dynamics. At the same time, time-aware methods require the choice of an observation timescale, yet we lack principled ways to determine it. We focus on the popular community detection problem which aims to partition a network's nodes into meaningful groups. We use a multi-layer quality function to show, on both synthetic and real datasets, that the observation timescale that leads to optimal communities is tightly related to the system's intrinsic aging timescale that can be inferred from the time-stamped network data. The use of temporal information leads to drastically different conclusions on the community structure of real information networks, which challenges the current understanding of the large-scale organization of growing networks. Our findings indicate that before attempting to assess structural patterns of evolving networks, it is vital to uncover the timescales of the dynamical processes that generated them.
\end{abstract}

\maketitle

\section{Introduction}
Many systems that are of interest for social science, information science, and data mining can be represented as complex networks that are not static but grow with time. For example, the global scientific output grows at a fast and accelerating pace, which results in growing citation networks of scientific papers that represent our accumulated scientific knowledge~\cite{price1965networks,bornmann2015growth}. The World Wide Web~\cite{huberman1999internet} and social contact networks~\cite{kumar2010structure,sekara2016fundamental} grow as well.
The presence of growth challenges traditional network analysis~\cite{wasserman1994social,newman2010networks} and makes it essential to develop and validate time-aware methods to achieve a solid understanding of the structure of these systems~\cite{holme2012temporal}.
In particular, extensive research has shown that the inclusion of temporal information into network analysis has a dramatic impact on long-studied problems such as community detection~\cite{tantipathananandh2007framework,rosvall2014memory,xu2016representing}, node ranking~\cite{lerman2010centrality,xu2016representing, liao2017ranking}, dynamics control~\cite{li2017fundamental}, and spreading phenomena~\cite{read2008dynamic,scholtes2014causality,holme2016temporal}.

This article focuses on one of the fundamental problems in network science, the detection of communities~\cite{fortunato2016community}, which has received enormous attention from diverse research areas, including physics~\cite{fortunato2016community}, computer science~\cite{schaeffer2007graph}, ecology~\cite{mariani2019nestedness}, neuroscience~\cite{lynn2019physics}, and social science~\cite{holland1983stochastic, wasserman1994social}, among others. While the problem is not uniquely defined~\cite{fortunato2016community}, it can be generically described as the problem of determining whether there exists a meaningful partition of the network nodes into groups of nodes. The problem (also known as clustering in computer science~\cite{schaeffer2007graph,mahmood2016using}) has a long history, and it has been traditionally addressed using structural, static network analysis.

A popular approach to community detection is to maximize a function, called \emph{modularity}, which quantifies how much the total number of intra-community edges deviates from its expected value under a null model that preserves the individual nodes' number of connections~\cite{newman2004finding}. Modularity has been studied from many viewpoints, and it is widely-recognized as a standard tool in network analysis~\cite{fortunato2016community}. Despite past research and the wide use of modularity optimization in a broad range of contexts, we still lack a systematic understanding of its behavior and performance in growing networks where time and aging phenomena are fundamental~\cite{dorogovtsev2000evolution,medo2011temporal,golosovsky2017growing}. Albeit modularity has been used in such systems in its original form~\cite{leicht2007large,chen2010community,stella2018bots}, the results can be expected to be suboptimal as modularity neglects the vital time information. A multi-layer form of modularity has been developed that can take into account network snapshots at various times~\cite{mucha2010community,granell2015benchmark}. However, when we wish to apply a multi-layer approach to identify relevant communities in growing networks, we face an impasse: Existing works assume layered input data~\cite{mucha2010community,granell2015benchmark,bazzi2016community,chai2016functional} and thus they do not consider the question of how to divide an arbitrary time-stamped network into layers. Addressing this question requires to choose an appropriate observation timescale, \ie, the temporal duration for each layer~\cite{krings2012effects, darst2016detection,sekara2016fundamental}. This choice is essential because different timescales might reveal substantially different community structures, which in turn might lead to different conclusions on the large-scale organization of the system.

In this work, we derive analytically a criterion to estimate when a time-aggregated, static view of a growing network ceases to be sufficient for effective community detection through standard modularity maximization. When this criterion is not met, the detected communities are strongly determined by node age and therefore in discordance with the network's actual community structure. We introduce the observation timescale $\tau_O$, divide the input network into subsequent layers of temporal duration $\tau_O$ each, construct a corresponding multi-layer modularity function, and use the resulting community detection method on diverse synthetic and real datasets. Remarkably, we find that the observation timescale $\tau_O^*$ that best uncovers the ground-truth communities in synthetic data is tightly related to the inherent aging timescale $\tau_S$ of the system's growth dynamics: $\tau_O^*\simeq\tau_S$. We use both synthetic and real data to show that different choices of the temporal resolution parameter lead to very different detected communities and conclusions on their statistical significance. Our results provide clear guidelines for data-driven calibration of multi-layer community detection techniques. Beyond the particular problem of community detection, the connection between the observation timescale $\tau_O$ used for structural analysis and the system's intrinsic timescale $\tau_S$ is relevant to the general problem of analyzing the structure and function of the broad variety of networks that evolve in time.

\section{Impact of network growth on modularity}
Before detailing the multi-layer community detection method, we start by demonstrating how temporal effects impair the ability of the traditional modularity maximization to uncover the community structure of growing networks with aging. Since aging is common for information networks where connections between the items are usually directed (such as citations in a scholarly citation network and followers in a social network), we use here the formalism of directed networks. A similar analysis is possible for undirected and bipartite networks.

\subsection{Modularity}
The classical Newman-Girvan modularity function~\cite{girvan2002community} has been adapted to quantify the quality of a partition into communities in directed networks~\cite{arenas2007size,malliaros2013clustering} as
\begin{equation}
\label{dir_modularity}
Q = \frac1{m}\sum_{i,j}\bigg(A_{ij} - \frac{k_i^{out} k_j^{in}}{m}\bigg)\delta(c_i,c_j)
\end{equation}
where $m$ is the number of network links, $A_{ij}$ is an element of the adjacency matrix which is $1$ if node $i$ points to node $j$ and zero otherwise, $k_i^{out}$ and $k_i^{in}$ are respectively the out- and in-degree of node $i$, $c_i$ denotes the community of node $i$, $\delta(c_i,c_j)$ is the Kronecker delta which is $1$ when $c_i=c_j$ and zero otherwise, and the summation indexes $i$ and $j$ run over all $N$ network nodes. Eq.~(\ref{dir_modularity}) is referred to as \emph{static modularity} from now on. The task is to find the network partition that maximizes static modularity which thus serves as the objective function. Among the various existing approaches to modularity optimization, the Louvain algorithm~\cite{blondel2008fast} is a popular choice.

The negative term in Eq.~(\ref{dir_modularity}) represents the expectation of $A_{ij}$ for a random network that has the same in- and out-degree sequence as the original network. However, such randomization is of limited use in growing networks where time plays an important role in the nodes' connection patterns~\cite{medo2011temporal,golosovsky2017growing}: It neglects the original network's temporal properties and, consequently, it can violate the network's fundamental temporal constraints. In often-studied citation networks, for example, it generates ``unphysical'' randomized networks where papers can also cite future papers~\cite{ren2018randomizing}. To demonstrate the problem of standard modularity, and later to assess its modification suitable for growing networks, we set up a simple network growth model based on the classical preferential attachment process with aging~\cite{medo2011temporal}.

\subsection{Model for growing networks with community structure}
\label{sec:model}

\begin{table}
\centering
\begin{ruledtabular}
\begin{tabular}{lll}
   Variable type & Variable & Meaning\\
\hline
 Model parameter & $n_0$ & Initial number of nodes\\
                 & $N$ & Final number of nodes\\
                 & $k^{out}$ & Outdegree of the introduced nodes\\
                 & $\Theta $ & Aging parameter\\
                 & $\mu$ & Community-mixing parameter\\
\hline
Network property & $m$ & Number of links\\
                 & $f_B$ & Fraction of inter-community links\\
                 & $\kavg$ & Average degree\\
\end{tabular}
\end{ruledtabular}
\caption{\textbf{Network model summary.} Adopted notation for the model parameters and the resulting network properties.}
\label{tab:notation}
\end{table}

The model assumes that each node belongs to one of two ground-truth communities and preferentially (to a degree that can be tuned in the model) links to other nodes in the same community. The ground-truth community of node $i$, $C_i$, is chosen at random. The model can be easily extended to a case with more communities of various relative size. There are initially $n_0$ nodes that are all assumed to appear at time $0$. The network then grows in time steps $t=1,\dots, N - n_0$. In each time step, one node is introduced in the network; the final number of nodes is thus $N$. Each introduced node creates $k^{out}$ outgoing links to the existing nodes. The probability that node $i$ points to an existing node $j$ is
\begin{equation}
\label{model}
P_{i\to j} = \frac{X_{ij}(k_j^{in} + 1)\ee^{-(i - j) / \Theta}}
{\sum_j X_{ij} (k_j^{in} + 1)\ee^{-(i - j) / \Theta}}
\end{equation}
where $k_j^{in}$ is the current indegree of node $j$, exponential aging controlled by the parameter $\Theta$ is assumed, and the general preference for links between nodes $i$ and $j$ is encoded in the term $X_{ij}$ which is described below. If node $j$ has been chosen by node $i$ before, the choice is repeated. In this way, there is at most one directed link between any two nodes in the network. Small $\Theta$ values result in ``short'' links that connect the newly introduced node with other recently introduced nodes. As $\Theta$ increases, aging slows down and becomes negligible when $\Theta\gg N$. We refer to Table~\ref{tab:notation} for a summary of the notation for all model parameters and network properties.

\begin{figure}
\centering
\includegraphics[scale=0.72]{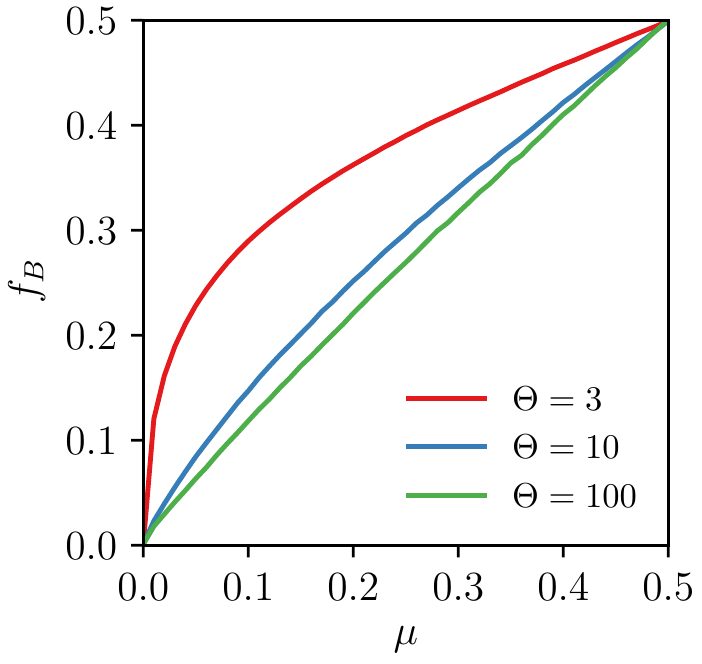}
\caption{\textbf{The relation between the community-mixing parameter $\mu$ and the resulting fraction of inter-community links $f_B$.} In the described network model, $f_B$ shows a non-linear yet monotonous dependence on $\mu$. As an illustration, we show here the fraction of links between communities $f_B$ as a function of $\mu$ for model data with $n_0=10$, $N=512$, $k^{out}=10$, and two ground-truth communities.}
\label{fig:f_BETWEEN}
\end{figure}

Note that Eq.~(\ref{model}) for simplicity omits the fitness term that is crucial to control the resulting network degree distribution~\cite{medo2011temporal}. We consider various model variants, including a variant with heterogeneous node fitness values, in Supplementary Material (SM). The community structure is introduced in the model by assuming
\begin{equation}
X_{ij} = \mu[1 - \delta(C_i,C_j)] + (1-\mu)\delta(C_i,C_j)
\end{equation}
where $C_i$ and $C_j$ are the ground-truth communities of $i$ and $j$, respectively. $X_{ij}$ is $1-\mu$ if nodes $i$ and $j$ are in the same ground-truth community and $\mu$ if they are not. As a result, the number of links between the communities grows with $\mu$. Other benchmark models for community detection, such as the Lancichinetti--Fortunato--Radicchi model~\cite{lancichinetti2008benchmark}, allow by construction to directly set the fraction of inter-community links in model networks. In our growth model, instead, $\mu$ only influences the preference for intra-community links in the network growth. The resulting fraction of inter-community links, $f_B$, therefore emerges by an interplay of this preference, the pool of available target nodes, and aging. Fig.~\ref{fig:f_BETWEEN} shows that $f_B$ grows with $\mu$ in a non-linear yet monotonous manner. In our numeric simulations, we achieve the desired $f_B$ values by choosing the appropriate $\mu(\Theta)$.

\begin{figure*}
\centering
\includegraphics[scale=0.67]{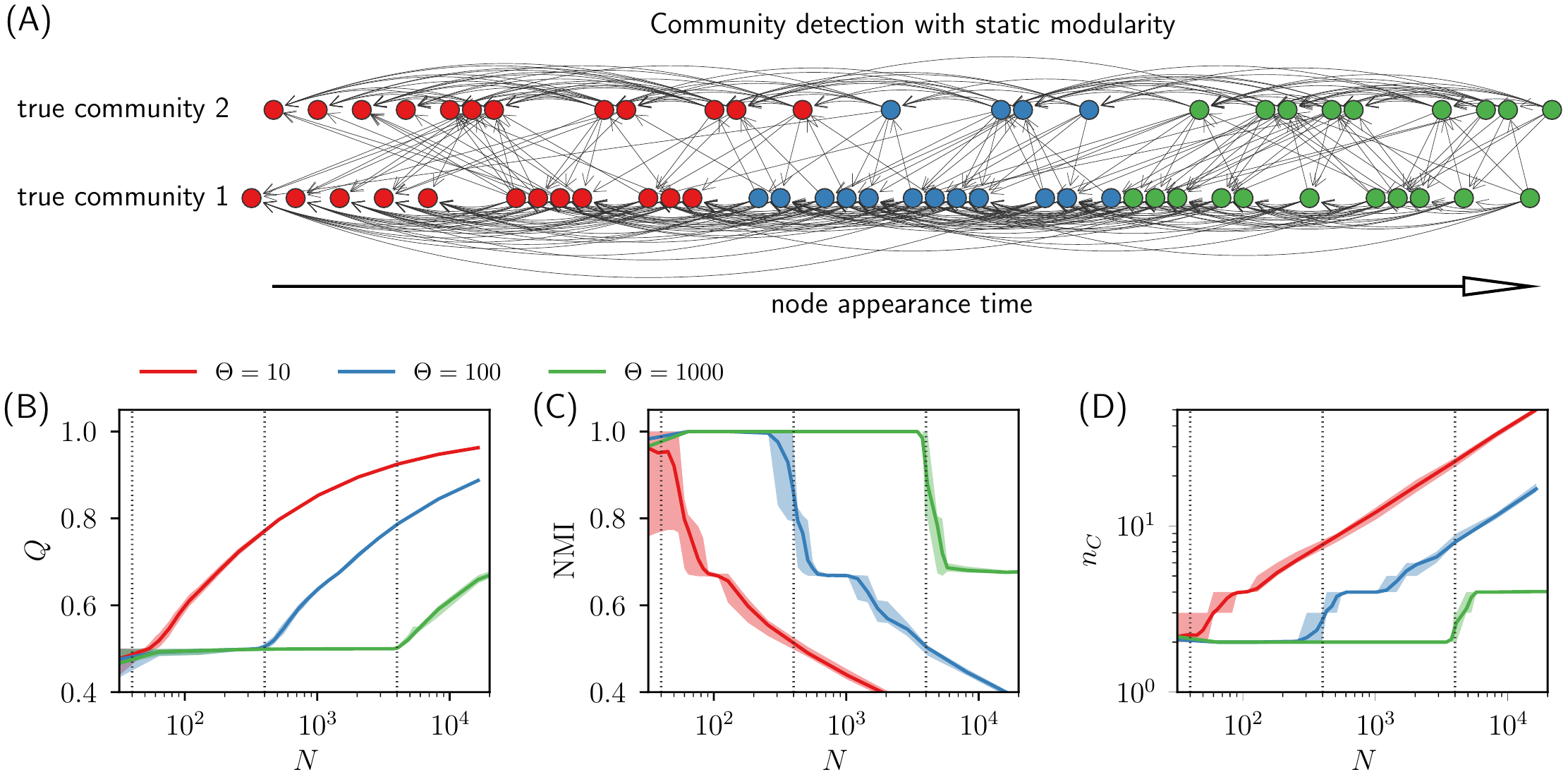}
\caption{\textbf{The breakdown of modularity in growing networks.} (A) Temporal confinement of communities identified by the standard static modularity in a model network with fast aging ($N=60$, $\Theta=3$, $k^{out}=10$, $\mu=0.05$; the resulting fraction of links between the communities is $f_B=0.24$). The horizontal position of nodes is given by their appearance time, the vertical position is given by their true community affiliation, and node color marks the community assigned by modularity maximization. We see that static modularity is essentially insensitive to the true communities and clusters the nodes by their appearance time.\\
(B--D) The behavior of static modularity on model data with different aging timescales (in all simulations $k^{out}=5$ and $\mu=0$, hence $f_B=0$; results are averaged over 100 model realizations and the shaded areas visualize the 10th-90th percentile range). Albeit the two ground-truth communities are perfectly separated in the model data, from some network size, modularity optimization yields inferior results with increasing modularity (panel B) yet decreasing NMI (panel C). The loss of NMI is due to the number of identified communities $n_C$ which is correctly two until some $N$ but then it starts to grow (panel D). The vertical lines mark the analytically estimated thresholds of $4\Theta$ for the three plotted $\Theta$ values.}
\label{fig:breakdown}
\end{figure*}

\subsection{Breakdown of static modularity in growing networks}
\label{analytical}
Fig.~\ref{fig:breakdown}A shows the result of modularity maximization in a toy model network. We see that, despite the two true communities being visually well separated, the result is markedly wrong as the nodes are essentially clustered by their appearance time. To understand why modularity fails to recognize the true communities, we focus now on the simple case where the two ground-truth communities are of similar size and perfectly separated (in the model, this is achieved by setting $\mu = 0$). For the correct partition into two communities that are identical with the ground-truth communities, the modularity contribution from true community $k$ is
\begin{equation}
Q_k = \frac1m\bigg(\sum_{i,j\TRU_k} A_{ij} - \frac{k_i^{out}k_j^{in}}m\bigg) =
\frac1m\bigg(\frac{m}{2} - \frac{(m/2)^2}{m}\bigg) = \frac14.
\end{equation}
The total modularity of the correct division is thus $1/2$. We now study the impact of dividing one correct community into two parts with sizes $N_1$ and $N_2$ ($N_1+N_2=N/2$). Modularity contribution from these two parts that cover true community $k$ is
\begin{equation}
Q_k'=\frac1m\bigg(\frac{m}2 - \Delta m - \frac{(\kavg N_1)^2}{m} - \frac{(\kavg N_2)^2}{m}\bigg)
\end{equation}
Here, $\Delta m$ represents the ``loss'' of links that are in the true community $k$ but do not contribute to $Q_k'$ because they run between the two parts that we now consider (\ie, they cross the partition boundary). The third and the fourth term represent the sum of expectation terms which goes over all $N_1^2$ pairs of nodes in the two parts into which $k$ is subdivided. The smallest sum of expectation terms is achieved when $N_1=N_2=N/4$ (\ie, the true community is divided into two parts of the same size) when we obtain
\begin{equation}
Q_k' = \frac1m\bigg(\frac{m}{2} - \Delta m - \frac{m}{8}\bigg) = \frac38 - \frac{\Delta m}{m}
\end{equation}
where we used $m = \kavg N$.

At this point, we can understand why a division of a single ground-truth community into more parts can increase modularity. If the average degree $\kavg$ is fixed, the number of links $m$ is proportional to the number of nodes $N$. At the same time, aging suppresses the formation of ``long'' links in the network and therefore implies an upper bound on the number of links between the two parts, $\Delta m$. As the network grows, the negative term $\Delta m/m$ therefore decreases and, thanks to a higher absolute term, $Q_k'$ thus eventually exceeds $Q_k$. At this point, modularity maximization results in dividing the true community $k$ into two parts. As the network grows further, the divisions continue and the number of identified communities grows (see Fig.~\ref{fig:breakdown}D).

We now estimate the network size at which the first division occurs. We do so assuming the exponential aging that we use in our numerical simulations. If we disregard preferential attachment, the probability that a link created under exponential aging targets a node introduced $n$ steps ago is approximately $\exp(-n/\Theta)/\Theta$ (we assume that $\Theta$ is large, so summation over individual $n$ values can be replaced with integration that eventually yields the normalization factor $\Theta$). The first node after the partition boundary must necessarily point all its outgoing links across the boundary. For a node $n$ steps after the partition boundary, the fraction of boundary-crossing links created by this note is approximately $\sum_{i=n}^{\infty} \exp(-i/\Theta) /\Theta\approx \exp(-n/\Theta)$. The total number of boundary-crossing links is then obtained by multiplying with $k^{out}$ and summing over all $n$ values. In the summation, each $n$ values carries the weight $1/2$ because there are two communities and we count the boundary-crossing links in only one of them. Hence
\begin{equation}
\overline{\Delta m} \approx \frac{k^{out}}{2}\sum_{n=1}^{\infty} \ee^{-n/\Theta}\approx\frac{\Theta k^{out}}2.
\end{equation}
When the initial number of nodes, $n_0$, is small, $k^{out}\approx\kavg$. The inequality $Q_k'>Q_k$ can be now solved for $N$, yielding
\begin{equation}
\label{criterion}
N > 4\Theta
\end{equation}
When this condition is met, network divisions into four (or more) parts are preferred to the correct division into two ground-truth communities.

Our analytically-derived criterion predicts accurately the breakdown of modularity in numeric experiments in the absence of inter-community links (Fig.~\ref{fig:breakdown}, panels B--D). In particular, when the criterion defined by Eq.~(\ref{criterion}) is met, the optimal modularity is larger than the value ($Q=0.5$) expected for a partition with two perfectly-separated communities (Fig.~\ref{fig:breakdown}B), the Normalized Mutual Information (see \ref{sec:metrics} for the definition) between detected and ground-truth communities is significantly smaller than one (Fig.~\ref{fig:breakdown}C), and the number of detected communities is larger than two (Fig.~\ref{fig:breakdown}D). In other words, in growing networks above some network size, modularity optimization fragments the ground-truth communities into smaller communities that are mostly determined by the nodes' age.

\section{Community detection in growing networks}
To resolve the limitations of static modularity, we propose the \emph{temporal modularity} quality function building on the recently-introduced Dynamic Configuration Model (DCM) for growing networks~\cite{ren2018randomizing} which proposes a way of randomizing time-stamped networks whilst approximately preserving the time evolution of each node's degree.

\subsection{Multi-layer modularity}
\label{sec:temp_mod}
To define the temporal modularity function, we divide the network's links by time into $L$ layers of an equal number of links; layers $l=1$ and $l=L$ contain the earliest and latest links, respectively. Possible ties (several links created at the same time) can be solved by ordering them at random. Since ties are scarce in the real networks analyzed here, their impact is marginal. The numbers of outgoing and incoming links established by node $i$ in layer $l$ are denoted as $\Delta k_{i, l}^{out}$ and $\Delta k_{j, l}^{in}$, respectively (note that $\sum_l \Delta k_{i, l}^{out}=k_i^{out}$ and $\sum_l \Delta k_{i, l}^{in}=k_i^{in}$). The total number of links created in layer $l$ is $m_l$ ($\sum_l m_l=m$). Note that requiring the layers to have the same number of links is just one of the possible choices. Other simple choices are layers of an equal number of newly introduced nodes, and layers of equal physical timespan (we discuss the latter in Sec.~\ref{sec:real}). While different real networks can, in principle, require different ways of constructing the layers, the current choice of an equal number of links has the advantage of producing layers of comparable statistical power.

\begin{figure*}
\centering
\includegraphics[scale=0.67]{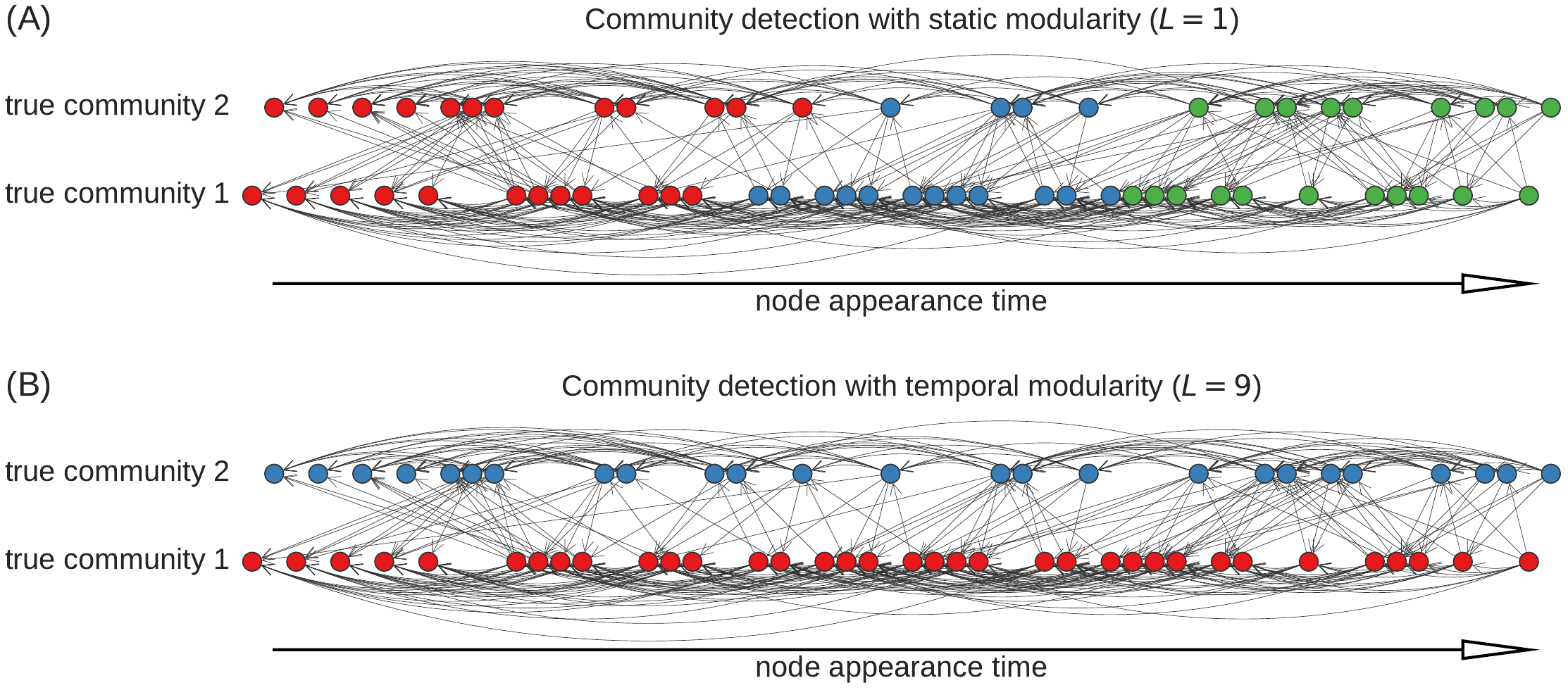}
\caption{\textbf{Comparing modularity and temporal modularity in a small network.} Partitions obtained by maximizing static directed modularity (top row) and the temporal modularity with 9 temporal layers (bottom row) in the model network from Fig.~\ref{fig:breakdown}A.
The choice $L=9$ follows from the observation timescale criterion described in Sec.~\ref{sec:optimal}.}
\label{fig:static_vs_temporal}
\end{figure*}

With the constraint of given degree increase sequences $\Delta k_{i, l}^{out}$ and $\Delta k_{j, l}^{in}$, the expectation of $A_{ij}$ can be written as
\begin{equation}
\avg{A_{ij}}_{DCM} = \sum_{l=1}^L \frac{\Delta k_{i,l}^{out} \Delta k_{j,l}^{in}}{m_l}
\end{equation}
where we assume that the connections between the nodes are random \emph{within each layer}. This expectation value can be used to replace the time-ignoring expectation value $k_i^{out} k_j^{in}/m$ in Eq.~(\ref{dir_modularity}) and obtain temporal modularity
\begin{equation}
\label{dir_temp_modularity}
Q_T(L) = \frac1m \sum_{i,j}\bigg(A_{ij} - \sum_{l=1}^L \frac{\Delta k_{i,l}^{out} \Delta k_{j,l}^{in}}{m_l}\bigg)\delta(c_i,c_j)
\end{equation}
where $L=1$ recovers static modularity defined by Eq.~(\ref{dir_modularity}). Similarly to static modularity, temporal modularity is zero when each node constitutes an independent community; the result of its maximization is thus non-trivial and depends on the network structure. Unlike previous work on communities in multilayer networks~\cite{mucha2010community}, we assume that node community affiliation does not change over time. If necessary, this assumption can be relaxed. We adapt the Louvain algorithm~\cite{blondel2008fast} to optimize temporal modularity (see~\ref{sec:algorithm} for details). In the toy example from Fig.~\ref{fig:breakdown}A, temporal modularity partitions the nodes correctly for any $L$ from $4$ to $20$ (see Fig.~\ref{fig:static_vs_temporal}).

Eq.~(\ref{dir_temp_modularity}) can be viewed as a special case of the previously introduced multi-layer modularity~\cite{mucha2010community,granell2015benchmark} where we constrain node affiliation to be fixed in time. While those studies assumed the layered structure of the data to be given, we obtained Eq.~(\ref{dir_temp_modularity}) by construction from continuously growing network data. As a result, the number of layers $L$ can be freely varied, and it is thus important to study how to choose it in practice. This problem was not investigated in previous studies on multi-layer generalizations of modularity~\cite{mucha2010community,granell2015benchmark} where the division of the network into layer was assumed to be given a priori, which is not the case for a generic time-stamped dataset.

\begin{table}
\centering
\begin{ruledtabular}
\begin{tabular}{ll}
Variable & Meaning\\
\hline
$\Theta$ & Aging parameter in the model\\
$L$ & Number of layers\\
$\tau_O:=N/L$ & Observation timescale (layer duration)\\
$e_n^{out}$ & Outgoing node of edge $n$\\
$e_n^{in}$ & Incoming node of edge $n$\\
$\tau_S$ & Detected link-based timescale (median of $\abs{e_n^{out}-e_n^{in}}$)\\
$\tau_J$ & Similarity-based timescale, extending Ref.~\cite{darst2016detection}\\
\end{tabular}
\end{ruledtabular}
\caption{\textbf{Timescales and related parameters.} Summary of the notation adopted in this paper. Note that the nodes are labeled by their order of appearance.}
\label{tab:notation_time}
\end{table}

One expects that the choice of $L$ is linked with the network's aging speed: While one layer may suffice when aging is slow or even absent, many layers are needed when aging is fast. To measure the aging speed, we measure the median span of links. Assuming that the nodes are labeled by the time of appearance and denoting the out-going and in-coming node of edge $n$ as $e_n^{out}$ and $e_n^{in}$, respectively, the system's median link span $\tau_S$ can be computed as the median of $\abs{e_n^{out}-e_n^{in}}$, that is
\begin{equation}
\tau_S:=\rm{median}\{\abs{e_n^{out}-e_n^{in}}\}.
\end{equation}
This timescale can be now compared with the average layer timespan
\begin{equation}
\tau_O:=N/L
\end{equation}
which defines the observation timescale of the new community detection method (see Table~\ref{tab:notation_time} for a summary of all the timescales and related variables considered in this article). If each layer covers time much longer than the aging timescale ($\tau_O\gg\tau_S$), the temporal effects ``average'' out and we can expect the results to be similar to those obtained with static modularity. By contrast, if layers are many and each of them contains only a handful of links ($\tau_O\ll\tau_S$), temporal modularity is expected to be hampered by statistical fluctuations. We thus expect temporal modularity to work best at an intermediate range of $\tau_O$ values; we will determine the optimal timescale below.

To evaluate community detection results on model data, we compute Normalized Mutual Information (NMI) between the detected partitions and the ground-truth communities. Motivated by the tendency of static modularity to produce temporally confined communities, we also compute the size-weighted average time span $\Omega$ of the detected communities which is related to the age difference between the oldest and the most recent nodes in each identified community. The higher the $\Omega$, the less temporally confined are the identified communities. The advantage of this metric is that it only concerns the properties of the detected communities and it does not involve any notion of ground-truth communities which cannot be uniquely defined in real data~\cite{peel2017ground}. We will thus use $\Omega$ to evaluate partitions in real networks, which will help us to establish a bridge between model-based observations and real data. Details for these two metrics and other metrics that further support our findings are described in \ref{sec:metrics}.

\begin{figure*}
\centering
\includegraphics[scale=0.72]{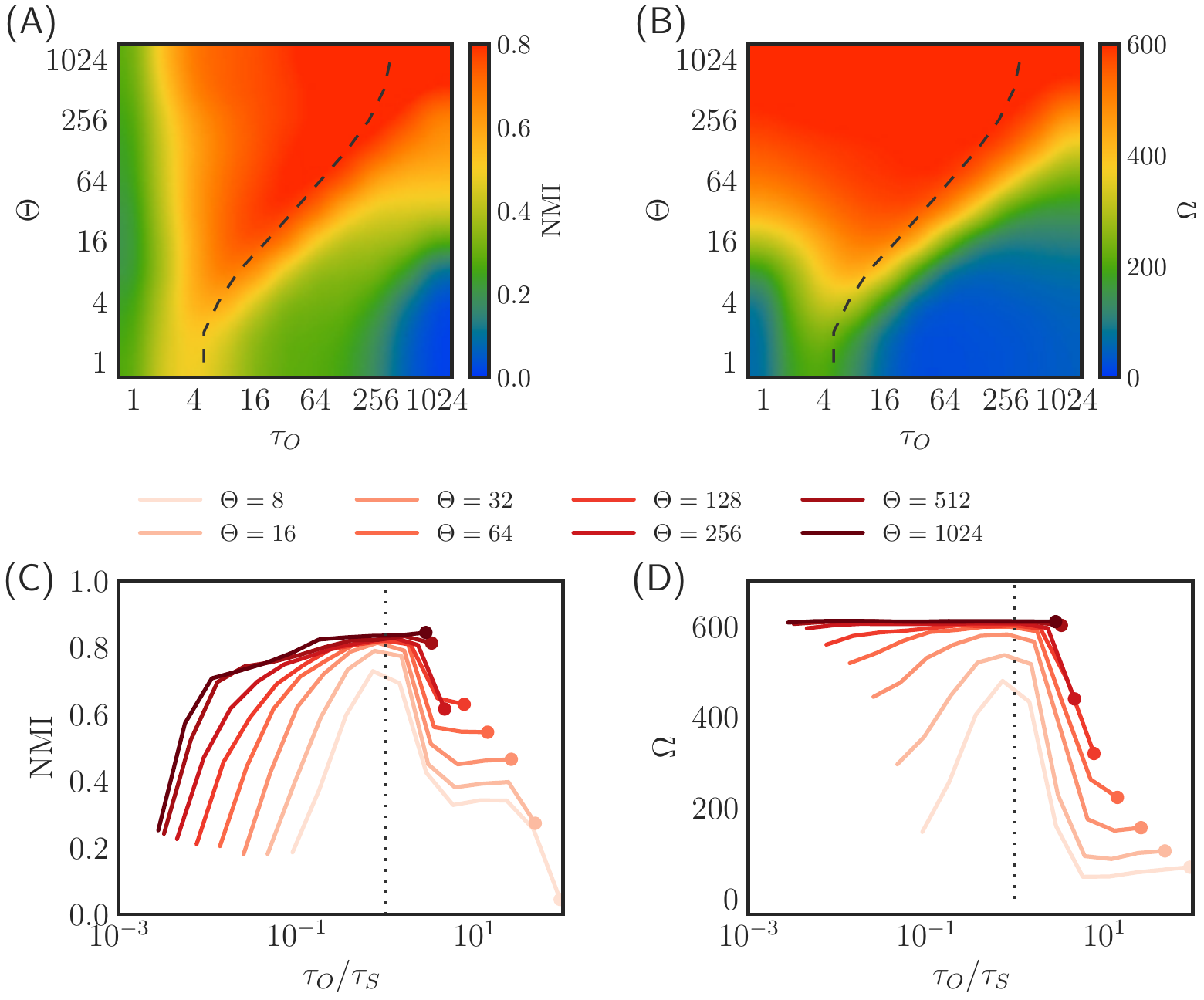}
\caption{\textbf{The optimal timescale for community detection in growing networks.} (A,B) Community detection results for model data as a function of the observation timescale $\tau_O$ and the model aging parameter $\Theta$: NMI (A) and the average community timespan (B). $\tau_O=1024$ corresponds to static modularity. The dashed lines mark the intrinsic system timescale $\tau_S$ corresponding to given $\Theta$. Model parameters: $N=1024$, $n_0=10$, $k^{out}=5$, $f_B=0.1$; results are averaged over 100 model realizations.\\
(C,D) To better appreciate the relation between $\tau_S$ and the optimal $\tau_O$, data from panels (A,B) are plotted here as a function of $\tau_O/\tau_S$. Full circles mark the right-most points of each curve which correspond to the results obtained with static modularity. These results show that the optimal NMI between the detected and ground-truth communities is achieved for $\tau_O\simeq\tau_S$; (2) the least time-confined communities (\ie, the communities with the largest $\Omega$ values) are observed for $\tau_O\simeq\tau_S$.}
\label{fig:model_data}
\end{figure*}

\subsection{The optimal timescale of temporal modularity}
\label{sec:optimal}
Results for model data with various aging timescales are shown in Fig.~\ref{fig:model_data}. While panels (A,B) explore the parameter plane $(\tau_O,\Theta)$, panels (C,D) use $\tau_O/\tau_S$ on the horizontal axis. As can be seen, both the NMI and $\Omega$ show a peak around $\tau_O/\tau_S\approx1$, in particular when aging is sufficiently fast ($\Theta\ll N$). The system's intrinsic timescale $\tau_S$ thus directly determines the optimal value of $\tau_O^*$ for the temporal modularity's layers. The loss of temporal modularity's efficiency when the observation timescale is too long compared to the system's timescale ($\tau_O/\tau_S\gtrsim 1$) is particularly fast. When $\tau_O/\tau_S\lesssim1$ (\ie, layers are shorter than the system timescale), we can observe performance plateaus that end when the number of nodes per layer becomes too small (in our simulations 10 nodes per layer or less) and the results become hampered by insufficient statistics. Panels (B,D) further show that the community time span, $\Omega$, can be indeed used to distinguish between the large-$\tau_O$ regime where communities are mainly determined by time and thus of limited time span (right side of the figure), an intermediate regime with ``long'' communities that are independent of time (hence they can reflect the network's structural information), and finally the noisy small-$\tau_O$ regime that again yields shorter sub-optimal communities.

\begin{figure*}
\centering
\includegraphics[scale=0.72]{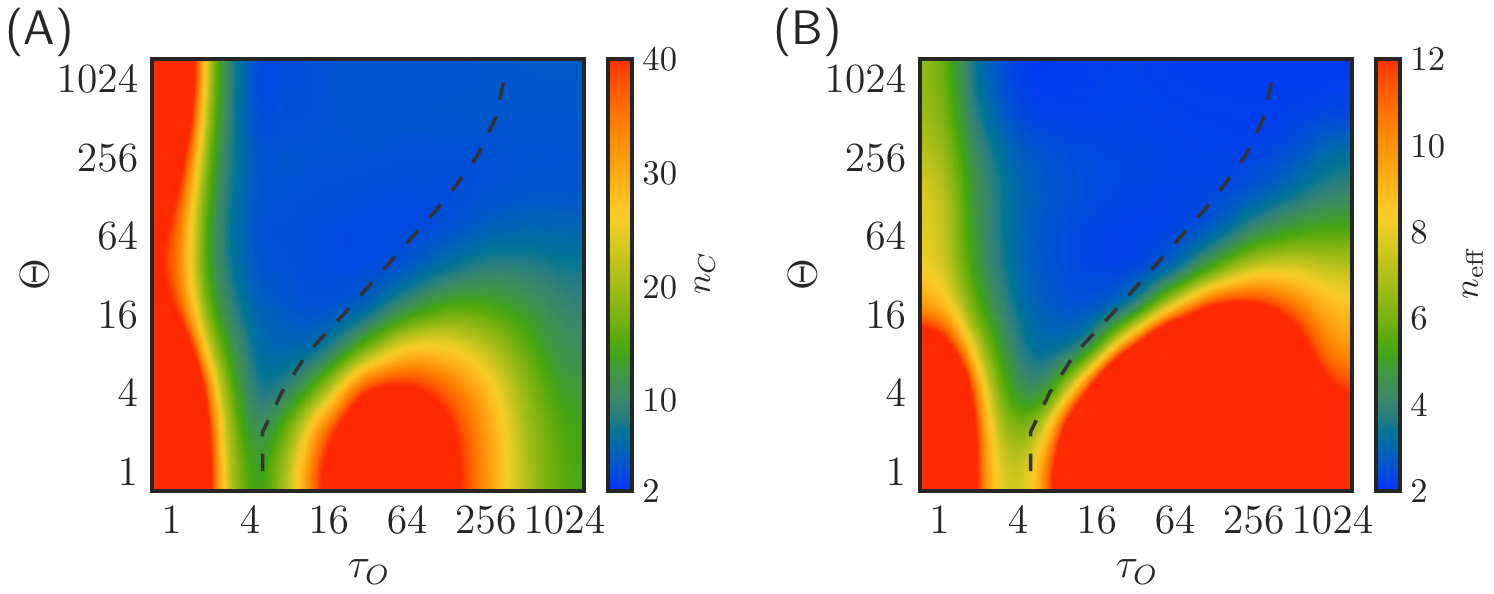}
\caption{\textbf{The number of detected communities in growing networks.} The average number of detected communities (A) and the average effective number of detected communities (B) for model data. Model parameters and other settings as in Fig.~\ref{fig:model_data}.}
\label{fig:heatmaps-nC_and_neff}
\end{figure*}

Additionally, Fig.~\ref{fig:heatmaps-nC_and_neff} shows results for two other evaluation metrics: the average number of detected communities and the average effective number of detected communities. We see that $\tau_O/\tau_S\approx 1$ leads to community divisions with fewer communities (both in absolute terms as well as measured by the effective number of communities $\neff$) than other choices of $\tau_O$. Results for other model settings and variants (see Figs.~S1--4 in SM) further confirm that choosing $\tau_O=\tau_S$ is optimal or nearly-optimal in many circumstances. The system's intrinsic timescale $\tau_S$ is thus an important connection between community-detection using temporal modularity and the system's intrinsic properties. Since $\tau_S$ is based on studying the temporal properties of individual links, in the following, we refer it as \emph{link-based timescale}.

Let us illustrate how the link-based timescale can be used to overcome the modularity breakdown illustrated in our initial example (Fig.~\ref{fig:breakdown}A). For this network, the link-based timescale is $\tau_S=7$. Setting the observation timescale $\tau_O$ to $7$ corresponds to choosing $L=N/\tau_O\approx 8.6$ layers in temporal modularity given by Eq.~(\ref{dir_temp_modularity}). In Fig.~\ref{fig:static_vs_temporal}, we rounded that up to 9 layers, leading to the perfect community detection result.

\subsection{Link-based and similarity-based timescale detection: A comparative analysis}
In the previous analysis, we have defined the system's timescale using the median link time span $\tau_S$. We now aim to compare this timescale detection criterion against an existing principled method for timescale detection in complex systems~\cite{darst2016detection}, which we refer to as \emph{similarity-based timescale detection}. The original method introduced in~\cite{darst2016detection} constructs iteratively layers $\mathcal{S}_n = [t_n, t_n + \Delta t_n)$ by maximizing the Jaccard similarity between the sets of events that occurred in pairs $(\mathcal{S}_n,\mathcal{S}_{n+1})$ of consecutive layers (the maximization is with respect to the layer duration $\Delta t_n$). The original method thus can, in principle, detect layers of heterogeneous lengths. However, as our synthetic networks feature a homogeneous timescale, we consider a variant of the method that aims to detect a single timescale. For given layer duration $\tau$, we compute the average Jaccard similarity between pairs of consecutive layers. The layer duration that leads to the maximal average similarity is selected.

The key element in the similarity-based timescale detection is the definition of an event. Since~\cite{darst2016detection} studies temporal networks, an event naturally is a temporal network link that occurs in layer $n$. In our case of a growing network, a link between two nodes appears at most once -- event sets composed of links would be therefore disjoint and their similarity would be zero for every layer duration $\tau$. We thus extend the original method and assume that an event is a node receiving a link in layer $n$ (as in~\cite{darst2016detection}, we consider unweighted events, \ie, it does not matter how many links a node has received). Denoting $\mathcal{S}_n$ the set of nodes that receive at least one link in layer $n$, the Jaccard similarity of layers $n$ and $n+1$ is $J=\abs{\mathcal{S}_n\cap\mathcal{S}_{n+1}} / \abs{\mathcal{S}_n\cup\mathcal{S}_{n+1}}$. The resulting timescale obtained by maximizing the average layer similarity is referred to as $\tau_J$.

\begin{figure*}
\centering
\includegraphics[scale=0.72]{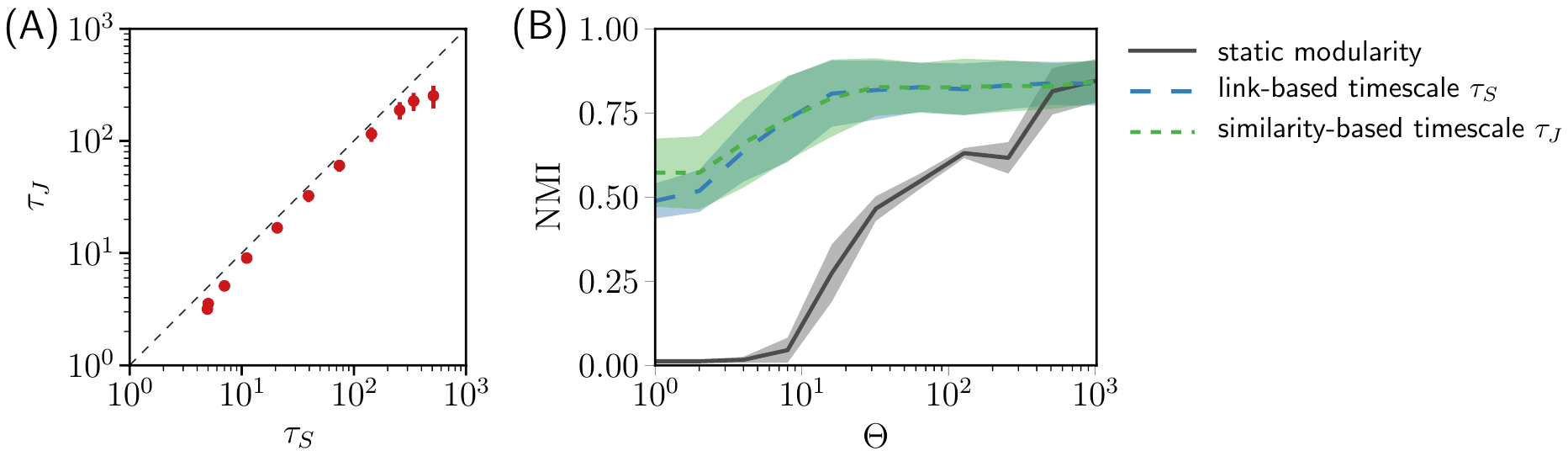}
\caption{\textbf{Comparing the performance of different timescale detection procedures.} Results are obtained on model data with parameters as in Fig.~\ref{fig:model_data} and the model aging parameter $\Theta$ varied in the range $[1, 1024]$. The lines and the shaded areas indicate, respectively, the means and the standard deviation values computed from 100 model realizations.}
\label{fig:timescales_comparison}
\end{figure*}

Our simulation results show that the two timescales, $\tau_S$ and $\tau_J$, have similar values across the whole range of the model aging parameter $\Theta$, and link-based timescales tend to be longer than the similarity-based timescales (Fig.~\ref{fig:timescales_comparison}A). More importantly, when the detected timescales are used to set the temporal modularity's layer duration, the detected communities are in much better agreement with the ground-truth communities than when static modularity is used. The two timescales yield similar NMI values across the whole range of $\Theta$ values except for the two smallest $\Theta$ values where the similarity-based timescales performs better than the link-based timescale. It has to be nevertheless noted that the link-based timescale is considerably simpler and amendable to analytical solutions than the similarity-based timescale.

\section{Significance analysis}
\label{sec:significance}
To assess the statistical significance of the detected communities, we compare the results obtained on the model networks with those on model networks where links are randomized within each layer. To this end, we implement the Dynamic Configuration Model (DCM, \cite{ren2018randomizing}) using the layer division described before (see Eq.~(\ref{dir_temp_modularity}) and its justification): all network links are sorted by the time of their appearance and divided into $L$ layers of equal size. In each layer $l$, the $\Delta k_{i, l}^{out}$ outgoing stubs of the nodes are matched with the $\Delta k_{j, l}^{in}$ incoming stubs of the nodes by choosing the candidate nodes preferentially by the number of remaining stubs. If the matched nodes correspond to an already existing link or a self-loop, the matching is repeated. In the end, there may be a small number of stubs that cannot be matched but this number is usually negligible. The number of layers in the DCM randomization is chosen depending on the system's timescale as $N/\tau_S$ rounded down.

Denoting the highest modularity values achieved on the original and DCM-randomized data as $Q_T(L)^{orig}$ and $Q_T(L)^{DCM}$, respectively, the significance of the detected communities can be evaluated by computing the $z$-score
\begin{equation}
z^{DCM} = \frac{Q_T(L)^{orig} - E[Q_T(L)^{DCM}]}{\sigma[Q_T(L)^{DCM}]}
\end{equation}
where $E[Q_T(L)^{DCM}]$ and $\sigma[Q_T(L)^{DCM}]$ denote the mean and the standard deviation of $Q_T(L)^{DCM}$ over multiple realizations of the DCM randomization (we use 100 realizations). When $z$ is small, the identified communities are not significant as their modularity is similar (or even worse) to the maximal modularity achievable in randomized data that are by construction free of any community structure. The higher the $z$ value, the more significant the network partition (commonly used significance thresholds for $z$-score are two and three).

We use bootstrap to estimate the uncertainty of the estimated $z$ scores. Denoting the estimated standard deviation of the resulting $z$-score as $\sigma[z]$, we can compare the significance of community partitions $A$ and $B$ by evaluating the normalized $z$-score difference
\begin{equation}
\delta z:=\frac{z_A-z_B}{\sqrt{\sigma[z_A]^2+\sigma[z_B]^2}}
\end{equation}
which compares the difference between the partitions' $z$-score with the combined uncertainty of the $z$-score estimates. Positive $\delta z$ means that partition $A$ is more significant than partition $B$. When $\delta z$ is large, the observed significance-difference is significant itself. In Fig.~\ref{fig:real_data}(C,F), we use the threshold of $3$ to decide between subsets where temporal modularity yields significantly less or significantly more significant communities than static modularity.

For the toy example from Fig.~\ref{fig:static_vs_temporal}, the described significance analysis shows that the result obtained at $L=9$ is much more significant than that obtained with static modularity (the $z$-scores are $15$ and $2.6$, respectively). Importantly, the usual significance analysis using the Configuration Model (which, similarly as static modularity, does not take time information into account and it is thus inappropriate in the current setting) deems the incorrect partition obtained with static modularity as highly significant (its $z$-score is $28$). This is an example of a static method that not only produces a misleading result but also confirms its statistical significance.

\begin{figure*}
\centering
\includegraphics[scale=0.72]{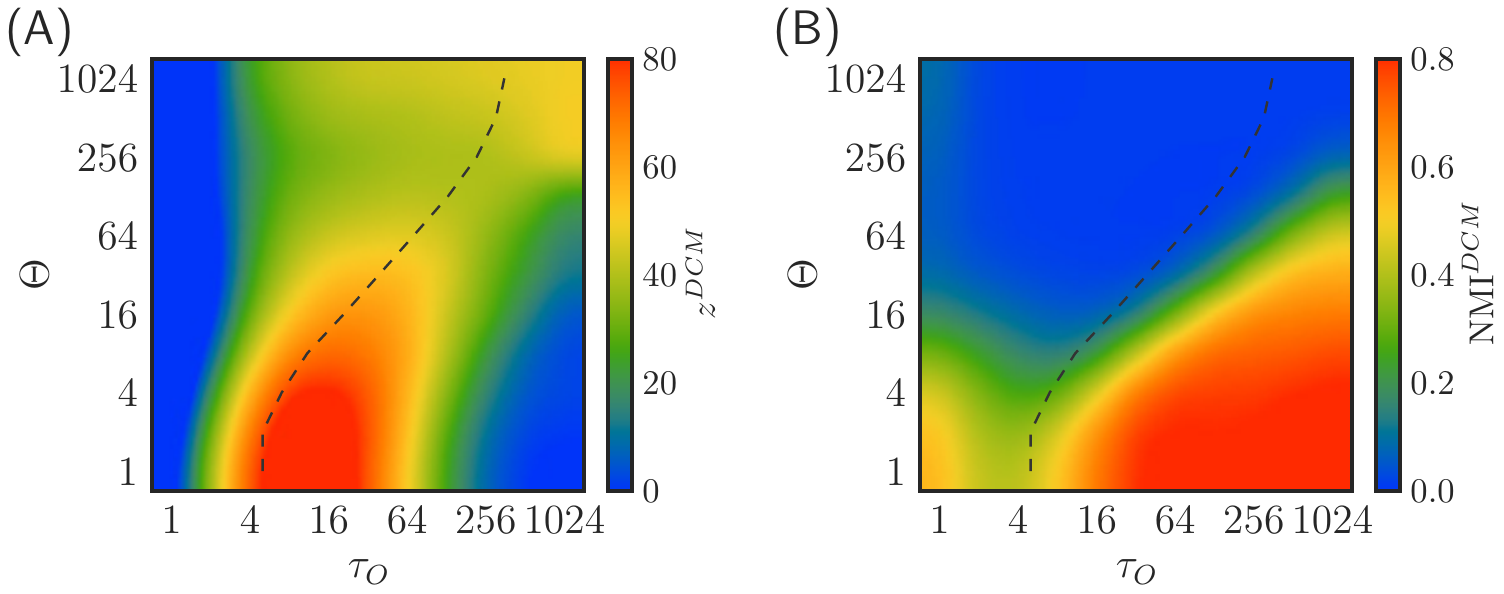}
\caption{\textbf{The statistical significance of the communities detected by the temporal modularity maximization.} We compare our previous results against randomized data: the $z$-score (A) and average NMI between the original and the communities detected in networks randomized with the Dynamic Configuration Model~\cite{ren2018randomizing} (B). Model parameters as in Fig.~\ref{fig:model_data}; results are averaged over 100 model realizations, each of them is randomized ten times.}
\label{fig:heatmaps-significance-DCM}
\end{figure*}

Significance analysis results for the basic model setting from Figure~\ref{fig:model_data} are shown in Figure~\ref{fig:heatmaps-significance-DCM}. We see that: (1) the detected communities are statistically significant when $\tau_O\approx\tau_S$, (2) the communities detected by static modularity are less significant or even not significant when aging is fast; this indicates that these communities are largely determined by node degree dynamics that is preserved by the DCM and not by higher-order structural patterns, (3) a too short observation timescale is also damaging to the significance of the detected communities. We compute also the average NMI between the originally identified communities and the communities identified in randomized data. Figure~\ref{fig:heatmaps-significance-DCM}(B) shows that the difference between the communities identified in the original and the randomized data is greatest when $\tau_O\approx\tau_S$, which confirms that temporal modularity is then most sensitive to the actual network structure.

\section{Implications for real networks}
\label{sec:real}
As $\tau_O\simeq\tau_S$ proves to be an optimal choice of the temporal modularity function in model networks, two natural questions arise: How different are the communities detected by temporal modularity optimization (with $\tau_O=\tau_S$) from those detected by static modularity maximization in real growing networks? Do the results obtained with temporal modularity call for a revision of our conclusions on the significance of the community structure of growing networks based on static modularity maximization?

To investigate the relation between temporal and static modularity maximization in real data, we use subsets of the news citation dataset that was used in~\cite{spitz2015breaking} to analyze the backbone of the citation network, and subsets of the American Physical Society (APS) citation data from years 1893--2013; the subsets correspond to specific newspapers and PACS codes, respectively (see \ref{sec:real_datasets} for details). We focus on three main properties: (1) the average community lifespan $\Omega$ of the detected communities -- we compare $\Omega_1$ achieved by static modularity with $\Omega_T$ achieved by temporal modularity; (2) the NMI between the communities detected by static and temporal modularity (since the true partition is unknown, we cannot directly evaluate the ``correctness'' of the obtained communities); (3) the normalized $z$-score difference between the partitions obtained with temporal and static modularity, respectively.

\begin{figure*}
\centering
\includegraphics[scale=0.66]{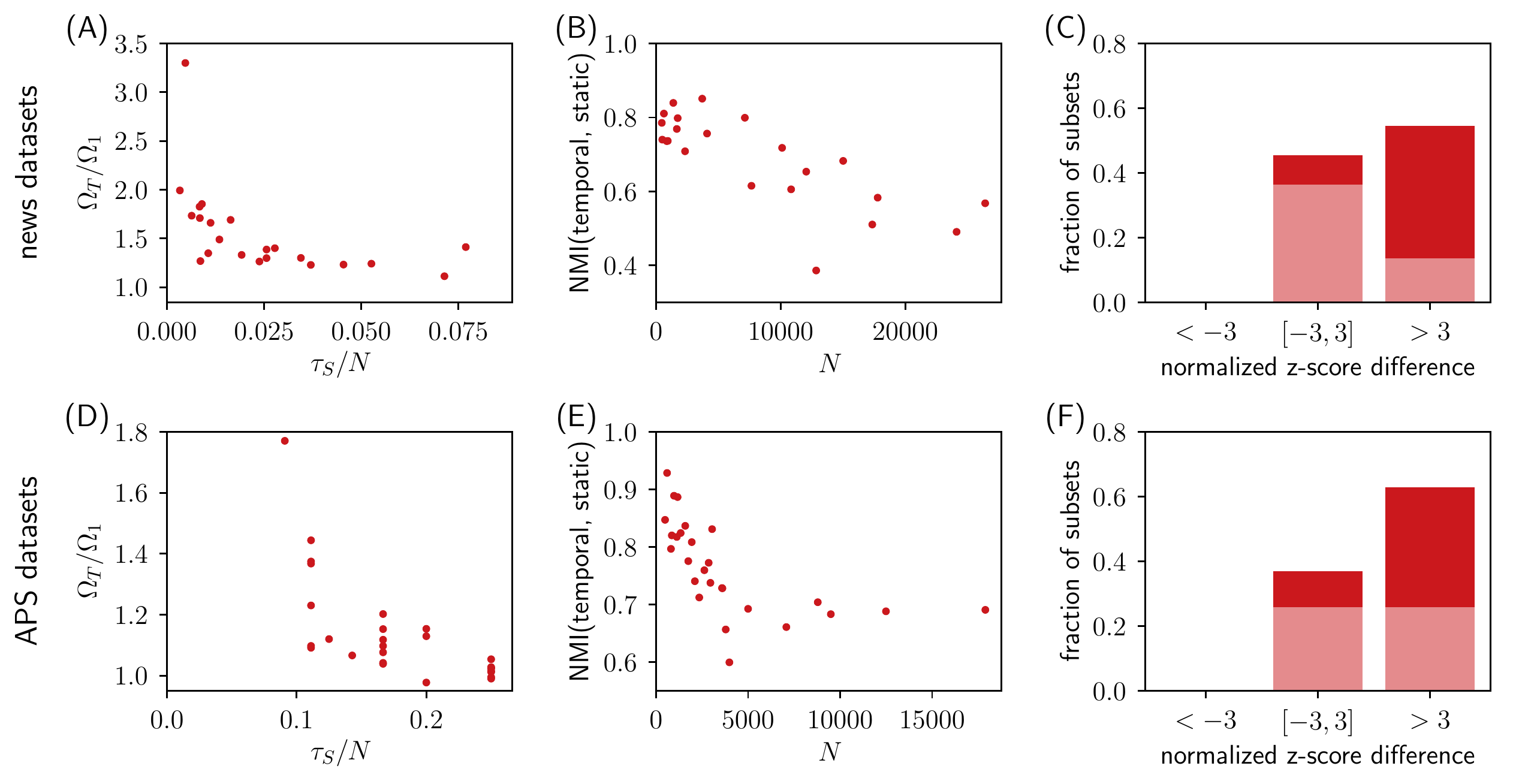}
\caption{\textbf{Implications of temporal effects for community detection in empirical growing networks.} We report the community detection results for the empirical News subsets (top) and the APS subsets (bottom); see \ref{sec:real_datasets} the data description. The columns show that, in order: (1) The average community time span $\Omega_T$ obtained with temporal modularity using $\tau_O=\tau_S$ is longer than the static modularity timespan $\Omega_1$, and the ratio $\Omega_T/\Omega_1$ tends to decrease with the relative aging speed $\tau_S/N$ (Spearman correlation $-0.77$ and $-0.78$ for the news and the APS data, respectively).
(2) NMI between communities obtained with temporal and static modularity tends to decrease with the number of nodes (Spearman correlation $-0.77$ and $-0.86$ for the news and the APS data, respectively).
(3) Communities obtained with temporal modularity tend to be more significant than those obtained with static modularity (the light and dark part of each column visualize the subsets with the number of nodes below and above the median for given datasets, respectively, indicating that the difference is even bigger for larger networks).}
\label{fig:real_data}
\end{figure*}

Figs.~\ref{fig:real_data}(A,D) show that the ratio $\Omega_T/\Omega_1$ is generally larger than one, confirming that the communities detected by temporal modularity have a longer time span than those detected by static modularity. Importantly, $\Omega_T/\Omega_1$ decreases with $\tau_S$: as aging becomes faster, the communities detected by temporal modularity become more ``stretched'' over time compared to those detected by static modularity. This is in qualitative agreement with our results on model data where the increase of $\Omega$ by temporal modularity grows with the aging speed (Fig.~\ref{fig:model_data}D). Figs.~\ref{fig:real_data}(B,E) show that the NMI between the communities by temporal modularity and those by static modularity is substantially smaller than one. Furthermore, the communities detected by the two methods tend to differ more for larger networks. Figs.~\ref{fig:real_data}(C,F) show that the communities by temporal modularity tend to be more statistically significant than the communities by static modularity. This indicates that properly including temporal information into the detection algorithm can substantially alter the conclusions on the significance of the community structure in growing networks. The analyzed real growing networks tend to be more ``temporally modular'' than modular: factoring out temporal patterns allows us to reveal a richer level of organization than possible with static modularity alone.

Recent literature~\cite{parolo2015attention} suggests that ``event time'' defined by the number of nodes characterizes the decay of attention in citation networks better than ``real time''. We have nevertheless considered an alternative to the equal layer size construction introduced in Sec.~\ref{sec:temp_mod} where real time is used instead to both define the layers as well as to measure the intrinsic system timescale $\tau_S$ which is then based on the median of real time differences between the appearance times of the out-going and the in-coming node. The obtained results qualitatively agree with those presented in Fig.~\ref{fig:real_data} without strong evidence in favor of either of the two temporal modularity constructions. Further research is necessary to map the settings where either of the two constructions is preferrable, and to understand what distinguishes them.

\section{Discussion}
The implications of our work are multi-fold. It has become increasingly clear~\cite{holme2012temporal,rosvall2014memory,scholtes2014causality} that to properly detect structural patterns in time-evolving systems, we need time-aware network analysis methods. While methods based on multi-layer representations of time-evolving networks are potentially powerful~\cite{mucha2010community,granell2015benchmark,chai2016functional}, they call for the fundamental question of how to choose the temporal resolution at which to look at the system~\cite{krings2012effects,darst2016detection}. Importantly, looking at a given system with different observation timescales can reveal different structural phenomena, such as different group organizations in social systems~\cite{sekara2016fundamental} and different behavioral patterns in communication networks~\cite{krings2012effects}, among others. Our work sets out to determine the optimal observation timescale $\tau_O^*$ for community detection based on a multi-layer generalization of the modularity function, which we called temporal modularity.

We found, both analytically and numerically, that modularity maximization yields unsatisfactory results in networks where aging decays sufficiently quickly. We found that the optimal observation timescale $\tau_O^*$ of the multi-layer modularity function is determined by the system aging timescale: $\tau_O^*\simeq\tau_S$. Different choices of $\tau_O$ lead indeed to sub-optimal performance in the reconstruction of ground-truth communities in model data, and to communities that are more strongly determined by node age in real data. The optimal timescale to look at the system is therefore close to the inherent timescale of the system's growth dynamics. This supports the idea that analyzing the structure of a time-evolving network requires to first understand the properties of the dynamical process that generated that network. Note that $\tau_S$ is not bound to the problem of community detection and characterizes the general aging pattern of a given network.

We proposed a simple procedure to detect communities in networks that organically grow over time. We suggested a novel metric, the link-based timescale $\tau_S$, to characterize the temporal patterns of a network, and demonstrated that this timescale should be used in the community detection process. When the aging timescale is short, the communities detected with our new framework are dramatically different from those detected with standard time-ignoring (static) approaches such as modularity maximization. Our comparison of the link-based timescale $\tau_S$ against similarity-based timescale $\tau_J$ based on previous literature~\cite{darst2016detection} reveals that the two timescales perform similarly for most parameter values, yet $\tau_S$ is conceptually simpler. In general, different timescale-detection methods can perform well for different tasks, and additional research is needed to study the performance of various timescales in different scenarios.

Our work paves the way for the search of community detection methods best suited to growing networks. Inspired by the equivalence between modularity and the long-known Stochastic Block Model~\cite{newman2016equivalence,pamfil2018relating}, an approach based on the dynamic stochastic block model~\cite{ghasemian2016detectability} can be compared against temporal modularity on our growing benchmark graphs. Besides, methods based on higher-order networks representations of temporal networks~\cite{rosvall2014memory} or consensus dynamics~\cite{lancichinetti2012consensus} can be tested within our framework. Various modifications (\eg, multiple communities of variable size) may improve our model to generate growing networks with aging and community structure, and the resulting models might be tested to explain the evolution of real growing networks~\cite{medo2014statistical}. While we focused here on growing networks, generalizing our results to temporal networks where links can appear or disappear over time~\cite{holme2015modern} is another open direction. Besides, cultural markets~\cite{easley2010networks} and E-commerce systems~\cite{zhou2010solving} can be represented with bipartite networks of users and products that grow over time. Adapting our approach to bipartite networks is thus a problem with many potential practical applications.

To conclude, we found that communities detected with temporal modularity are statistically more significant than those detected with standard modularity in the majority of analyzed empirical growing networks, which suggests that including temporal information into the community detection algorithm can unveil a richer large-scale organization than that uncovered by static methods. This calls for a note of caution on the use of the popular modularity maximization and, more generally, static community detection algorithms. Our findings demonstrate that if we analyze a time-evolving system, the communities detected by modularity maximization might be strongly influenced by the age of the nodes. A better way to detect communities in time-evolving networks is to first measure the typical timescale of the dynamic mechanisms that generated that network, and then exploit this information to analyze the system's structural patterns at that timescale.

\begin{acknowledgments}
MM, MSM and YCZ conceived the original idea. MM and MSM designed the study. AZ and MM acquired the data. MM and AZ performed numeric simulations. MM and MSM analyzed the results. MM and MSM wrote the manuscript.\\
All datasets necessary to reproduce the results in this paper will be publicly available.\\
This work was supported by the Swiss National Science Foundation Grant No. 200020-156188. M. S. M. acknowledges financial support from the University of Zurich through the URPP Social Networks. A. Z. acknowledges the support from the National Natural Science Foundation of China (grant No.~61603046).
\end{acknowledgments}

\appendix

\section{Community division evaluation metrics}
\label{sec:metrics}
\emph{Normalized Mutual Information} (NMI) is a standard evaluation metric in community detection research~\cite{fortunato2010community,fortunato2016community}. Denoting the sets of nodes comprising the detected communities as $\mathcal{C}_1,\dots,\mathcal{C}_D$ and the sets of nodes comprising the ground-truth communities as $\mathcal{G}_1,\dots,\mathcal{G}_T$, the normalized mutual information between the two community partitions is computed as
\begin{equation}
NMI =  \frac{-2\sum_{i=1}^D\sum_{j=1}^T\abs{\CF_i\cap\GT_j}\log\frac{N\abs{\CF_i\cap\GT_j}}{\abs{\CF_i}\abs{\GT_j}}}{\sum_{i=1}^D\abs{\CF_i}\ln\frac{\abs{\CF_i}}{N}+
\sum_{j=1}^T\abs{\GT_j}\ln\frac{\abs{\GT_j}}{N}}
\end{equation}
where $N$ is the total number of nodes. The terms that are of the kind $0\ln0$ are ignored in the summations.

Average community time span ($\Omega$) is introduced to measure the time confinement of communities. For a detected community $k$, we compute the 80th and 20th percentile of node IDs in the community, and define the community time span $\Omega_k$ as the difference between the two values. (The difference between the maximal and minimal node ID in the community would be more prone to outlier nodes.) The overall average time span is computed as a weighted mean of $\Omega_k$ with weights given by the community sizes
\begin{equation}
\Omega = \frac{\sum_{k=1}^D\abs{C_k}\Omega_k}{\sum_{k=1}^D\abs{C_k}}.
\end{equation}
Based on Fig.~1 in the main text and the accompanying discussion, we hypothesize that in systems with fast aging, optimization of static modularity leads to time-constrained communities with time span that is comparatively smaller than the time span of communities identified with temporal modularity.

For a given network partition one can evaluate the \emph{number of detected communities}, $n_C$. Since the distribution of community sizes can be uneven, we also compute the \emph{effective number of detected communities} (also known as the inverse Herfindahl index)
\begin{equation}
\neff = \frac{(\sum_k \abs{\COM_k})^2}{\sum_k \abs{\COM_k}^2}
\end{equation}
where $\abs{\COM_k}$ is the size of detected community $k$. When one community contains almost all network nodes, $\neff\to1$. When all $D$ detected communities have the same size, $\neff=D$.

\section{Algorithm for optimizing temporal modularity}
\label{sec:algorithm}
We maximize the temporal modularity defined by Eq.~(\ref{dir_temp_modularity}) by following the steps used by the Louvain algorithm that was originally proposed to maximize the standard Newman-Girvan modularity for undirected networks~\cite{blondel2008fast}. This algorithm is a ``greedy'' optimization algorithm which means that only configuration changes that increase the objective function are accepted. The algorithm proceeds as follows. Each node is initially in its own community ($c_i=i$). In the first step, individual nodes move from one community to another. In particular, we choose a node at random and search for the community, which yields the largest increase of $Q_T(L)$ if the node joins it. After moving the node to the best community, or after determining that there is no modularity-increasing move, we proceed with another node chosen at random. When no single node can be moved, optimization step one ends. Note that if a node that moves from a community is the last node of this community, the community effectively disappears. In this way, the number of communities progressively decreases during the first optimization step.

The second optimization step is similar but instead of individual nodes, we probe merging of entire communities. In particular, we choose a community at random and search for the community that yields the largest increase of $Q_T(L)$ upon merging with the chosen community. When no pair of communities can be merged, optimization step two ends and the final partition of nodes is reported.

To speed up the computation, the inner sum in Eq.~(\ref{dir_temp_modularity}), $\sum_{l=1}^L \Delta k_{i,l}^{out} \Delta k_{j,l}^{in} / m_l$, can be precomputed and stored in memory. Since the optimization algorithm contains randomness (we probe the nodes and communities, respectively, in random order), it is possible to run the algorithm several times and output the solution that yields the highest value of $Q_T(L)$. In our simulations, we always use 10 independent algorithm runs.

\section{Real datasets}
\label{sec:real_datasets}
To support our findings, we use two different real datasets that can be represented with growing directed monopartite networks: a news dataset~\cite{spitz2015breaking} and a citation dataset of papers published by the American Physical Society (APS).

\paragraph{News data.} The news citation dataset was used to analyze the backbone of the citation network and its impact on network centrality metrics~\cite{spitz2015breaking}. The dataset consists of news published by various outlets (newspapers and televisions) and citations among them. Since most citations are among the news published by the same outlet ($91\%$), we treat articles published by individual outlets as independent datasets. For each subset, self-loops and nodes that do not belong in the giant component are discarded. Only subsets with 500 edges or more are included in the further analysis (there are 22 of them). Table~S1 in SM summarizes the basic properties of the analyzed news subsets.

\paragraph{APS dataset.} We use the APS citation data that were obtained on our request from \url{https://journals.aps.org/datasets}. Our dataset contains all papers published by the APS from 1893 until December 2013 and the links among them. Importantly, paper metadata contains paper PACS codes which allow us to construct subsets of the original dataset in a controlled way (see \url{https://journals.aps.org/PACS} for details on the Physics and Astronomy Classification Scheme, PACS). The original data comprise 539,974 papers and 5,992,897 citations among them. Majority of the papers have some PACS codes assigned ($404,999$ out of $539,974$; most of the papers without PACS codes were published before 1977 when the PACS classification scheme was introduced). The PACS codes have a three-level hierarchy (for example, code ``89.75.-k'' represents ``Complex systems''). To construct subsets, we use the two top levels (for example, ``89.75.*'') which results in larger subsets than when complete PACS codes are used. Every paper that has a given two-level PACS code is considered as part of the subset. For each subset, self-loops and nodes that do not belong in the giant component are discarded. To create the subsets, we chose two-level PACS codes whose aging speed $N/\tau_S$ covers a broad range of values. Only subsets with at least 1000 edges are used for further analysis (there are 27 of them). Table~S2 in SM summarizes the basic properties of the analyzed APS subsets.

\clearpage

\onecolumngrid
\renewcommand\thefigure{S\arabic{figure}}
\setcounter{figure}{0}

\renewcommand\thetable{S\arabic{table}}
\setcounter{table}{0}

\begin{center}
\LARGE Supplementary Material
\end{center}

\section*{Additional results on model data}
\begin{figure}[b!]
\figSI{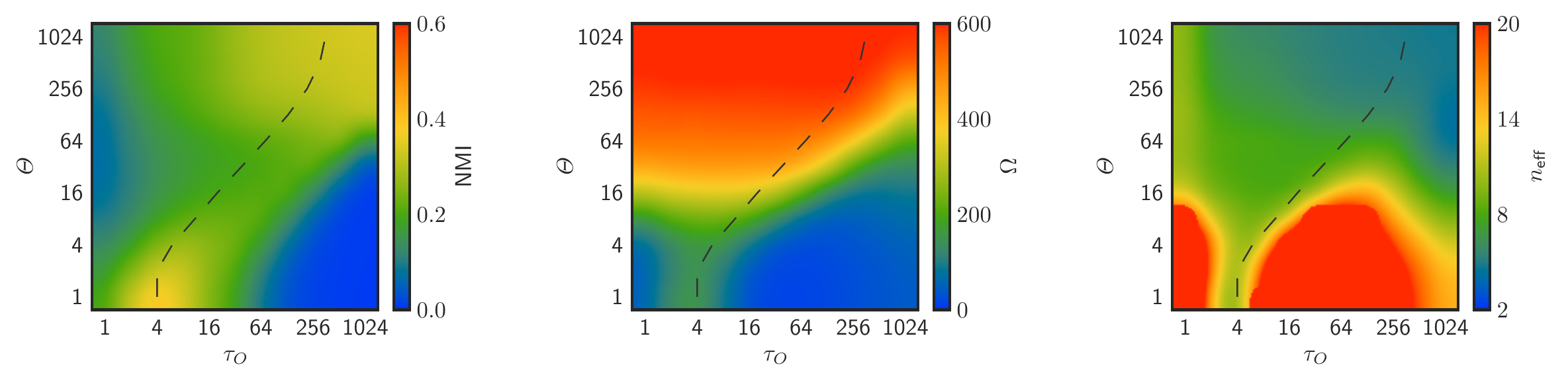}
\caption{As Fig.~4 in the main text but $f_B=0.2$ (\ie, the two ground-truth communities are more interconnected).}
\label{fig:heatmaps_noise}
\end{figure}

Results for all evaluated metrics in the parameter plane $(\tau_O,\Theta)$ are shown here as heatmaps. This complements Figure~2 in the main text which shows NMI and $\Omega$ for one model setting. Figure~4 in the main text shows the basic setting $N=1024$, $n_0=10$, $k^{out}=5$ and $f_B=0.1$. In addition to the observations mentioned in the main text, we see that choosing $\tau_O\approx\tau_S$ yields the smallest number of detected communities which is yet another positive contribution of temporal modularity. The agreement between the optimal $\tau_O$ and the system intrinsic timescale $\tau_S$ holds also when the communities are more interconnected (see Figure~\ref{fig:heatmaps_noise}). Naturally, the results of community detection are then worse in comparison with denser networks or those whose community structure is less noisy.

\begin{figure}[b!]
\figSI{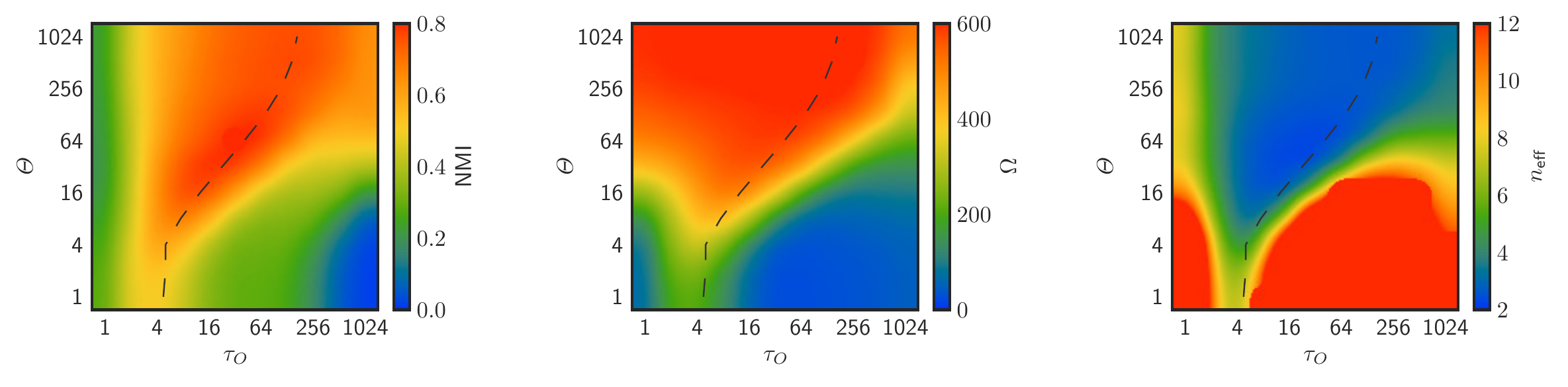}
\caption{As Fig.~4 in the main text but preferential attachment has been removed from the model; the attractiveness of nodes to new links is thus determined solely by their age.}
\label{fig:model1}
\end{figure}

\begin{figure}
\figSI{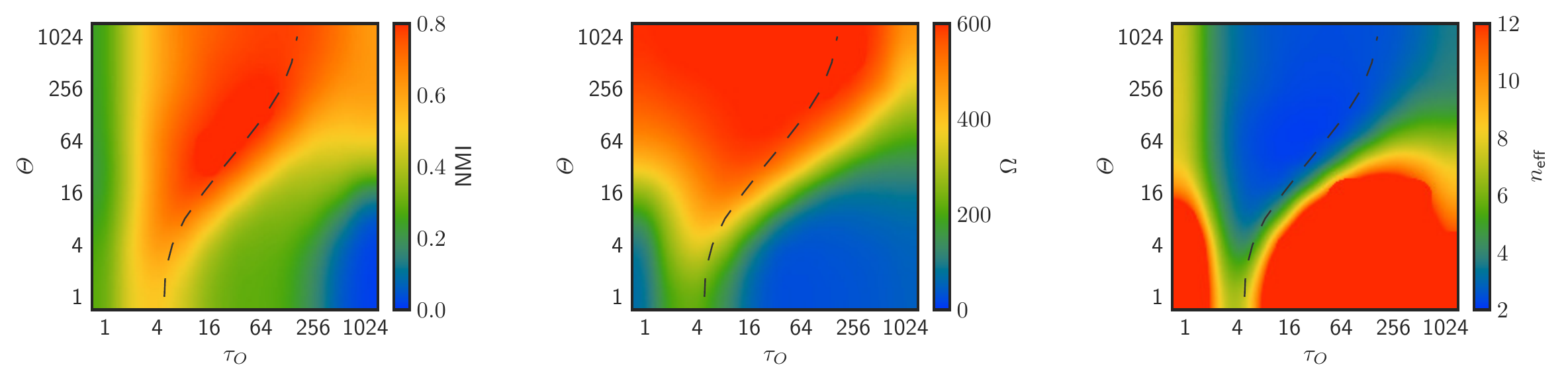}
\caption{As Fig.~4 in the main text but preferential attachment has been removed from the model and nodes are assigned fitness that is exponentially distributed.}
\label{fig:model2}
\end{figure}

\begin{figure}
\figSI{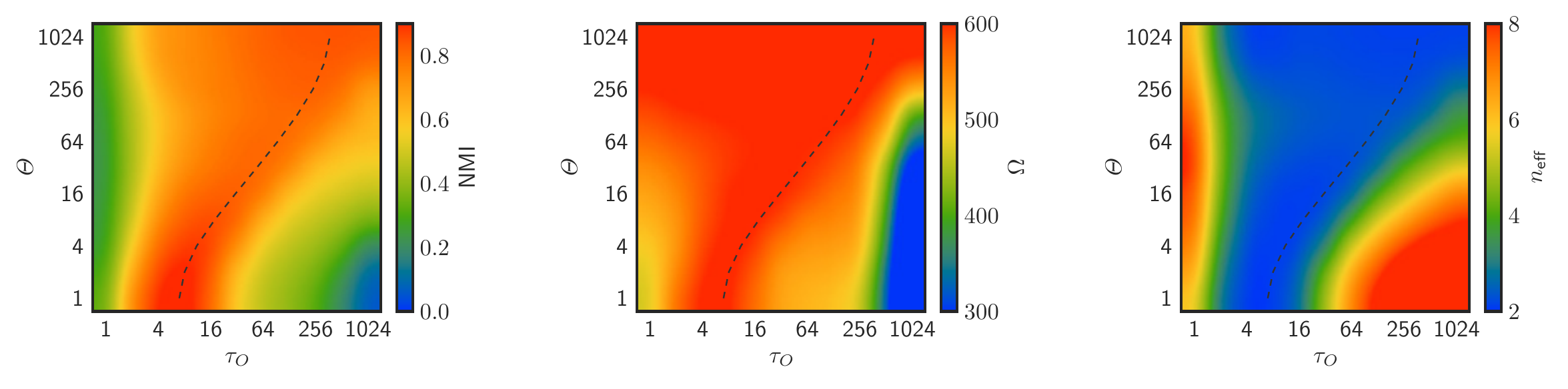}
\caption{As Fig.~4 in the main text but exponential aging has been replaced with power-law aging.}
\label{fig:model3}
\end{figure}

The basic model that we use for all simulations in the main text and all results above is based on preferential attachment and aging; differently from~\cite{medo2011temporal}, for simplicity there is no node fitness. Figures~\ref{fig:model1}, \ref{fig:model2} and \ref{fig:model3} show results for model data with substantial variations to the model that are achieved by modifying Eq.~(2) in the main text:
\begin{enumerate}
\item Model with aging, no preferential attachment, and no fitness: $P_{i\to j} \sim X_{ij}\ee^{-(i - j) / \Theta}$,
\item Model with aging and exponentially distributed node fitness, no preferential attachment: $P_{i\to j} \sim X_{ij} \eta_j \ee^{-(i - j) / \Theta}$ where node fitness values $\eta$ are exponentially distributed in the range $[1, \infty)$,
\item Same as the original model but exponential aging is replaced with power-law aging: $P_{i\to j} \sim X_{ij}(k_j^{in} + 1) / (1 + [(i - j) / \Theta]^2)$.
\end{enumerate}
In all three cases, the community structure-inducing term $X_{ij}$ remains the same as defined by Eq.~(3) in the main text. As can be seen in Figures~\ref{fig:model1}--\ref{fig:model3}, these modifications do not alter any of our main results.

\begin{table}
\centering
\begin{ruledtabular}
\begin{tabular}{rrrrr}
News outlet & $N$ & $m$ & $\mu$ & $N/\tau_S$\\
\hline
Guardian & 24122 & 45327 & 1.88 & 111.2\\
Die Welt & 26417 & 37610 & 1.42 & 118.5\\
Die Zeit & 17788 & 31381 & 1.76 & 157.4\\
Washington Post & 17355 & 29092 & 1.68 & 39.3\\
CBS & 12852 & 22953 & 1.79 & 210.7\\
Der Spiegel & 15018 & 21413 & 1.43 & 300.4\\
Los Angeles Times & 10116 & 20220 & 2.00 & 12.6\\
Independent & 12054 & 17510 & 1.45 & 88.6\\
Telegraph & 10840 & 17091 & 1.58 & 73.7\\
New York Times & 7656 & 12500 & 1.63 & 52.1\\
International Business Times & 7127 & 10069 & 1.41 & 41.7\\
Basler Zeitung & 3704 & 6798 & 1.84 & 39.0\\
Neue Zürcher Zeitung & 4093 & 5832 & 1.42 & 14.2\\
Toronto Star & 2327 & 3590 & 1.54 & 61.2\\
Sky News & 1667 & 2407 & 1.44 & 119.1\\
BBC & 1736 & 2288 & 1.32 & 115.7\\
Al Jazeera & 1384 & 1818 & 1.31 & 28.8\\
Süddeutsche Zeitung & 942 & 1171 & 1.24 & 94.2\\
Canadian Broadcasting Corporation & 854 & 1012 & 1.19 & 22.5\\
United Press International & 629 & 950 & 1.51 & 27.3\\
New Yorker & 500 & 722 & 1.44 & 35.7\\
Atlantic & 460 & 646 & 1.40 & 19.2\\
\end{tabular}
\end{ruledtabular}
\caption{Basic characteristics of the datasets corresponding to various news outlets in the dataset from~\cite{spitz2015breaking}; only the giant component is kept, subsets with less than 500 edges in the giant component are discarded. The shown characteristics: the number of nodes ($N$), the number of edges ($m$), mean degree ($\mu$), and the ratio $N/\tau_S$ which, as we argue, can be used to determine a suitable number of layers for temporal modularity.}
\label{tab:news}
\end{table}

\begin{table}
\centering
\begin{ruledtabular}
\begin{tabular}{rrrrr}
PACS code & $N$ & $m$ & $\mu$ & $N/\tau_S$\\
\hline
42.50-* & 17890 & 147261 & 8.23 & 6.3\\
03.67-* & 12491 & 122464 & 9.80 & 4.3\\
03.75-* & 8786 & 116448 & 13.25 & 5.1\\
98.80-* & 9496 & 97721 & 10.29 & 6.8\\
74.70-* & 7078 & 51389 & 7.26 & 10.7\\
14.60-* & 3983 & 37751 & 9.48 & 8.7\\
25.75-* & 3604 & 34646 & 9.61 & 6.4\\
74.60-* & 3785 & 31123 & 8.22 & 4.9\\
71.27-* & 4999 & 23603 & 4.72 & 5.7\\
04.25-* & 1939 & 23201 & 11.97 & 5.8\\
04.30-* & 2109 & 22505 & 10.67 & 5.8\\
95.35-* & 2347 & 19303 & 8.22 & 6.3\\
61.30-* & 3584 & 19261 & 5.37 & 5.9\\
72.25-* & 2953 & 16174 & 5.48 & 4.4\\
61.50-* & 3044 & 9746 & 3.20 & 9.5\\
45.70-* & 1750 & 8741 & 4.99 & 4.2\\
68.37-* & 2858 & 8169 & 2.86 & 4.2\\
72.80-* & 2624 & 6966 & 2.65 & 9.3\\
75.75-* & 1585 & 5702 & 3.60 & 4.2\\
87.16-* & 1337 & 4007 & 3.00 & 3.9\\
03.70-* & 810 & 3302 & 4.08 & 7.9\\
81.15-* & 1169 & 3290 & 2.81 & 5.1\\
52.27-* & 855 & 3279 & 3.84 & 4.1\\
68.65-* & 1126 & 2692 & 2.39 & 8.5\\
33.20-* & 976 & 2322 & 2.38 & 8.8\\
61.70-* & 598 & 1403 & 2.35 & 3.8\\
36.20-* & 489 & 1053 & 2.15 & 9.4\\
\end{tabular}
\end{ruledtabular}
\caption{Basic characteristics of the datasets corresponding various top-two-level PACS codes in the APS citation data until December 2013; only the giant component is kept, subsets with less than 1000 edges in the giant component are discarded. The shown characteristics: the number of nodes ($N$), the number of edges ($m$), mean degree ($\mu$), and the ratio $N/\tau_S$ which, as we argue, can be used to determine a suitable number of layers for temporal modularity.}
\label{tab:APS}
\end{table}


\begin{thebibliography}{10}
\bibitem{price1965networks}
Derek~J De~Solla~Price.
\newblock Networks of scientific papers.
\newblock {\em Science}, pages 510--515, 1965.

\bibitem{bornmann2015growth}
Lutz Bornmann and R{\"u}diger Mutz.
\newblock Growth rates of modern science: A bibliometric analysis based on the
  number of publications and cited references.
\newblock {\em Journal of the Association for Information Science and
  Technology}, 66(11):2215--2222, 2015.

\bibitem{huberman1999internet}
Bernardo~A Huberman and Lada~A Adamic.
\newblock Internet: growth dynamics of the world-wide web.
\newblock {\em Nature}, 401(6749):131, 1999.

\bibitem{kumar2010structure}
Ravi Kumar, Jasmine Novak, and Andrew Tomkins.
\newblock Structure and evolution of online social networks.
\newblock In {\em Link mining: models, algorithms, and applications}, pages
  337--357. Springer, 2010.

\bibitem{sekara2016fundamental}
Vedran Sekara, Arkadiusz Stopczynski, and Sune Lehmann.
\newblock Fundamental structures of dynamic social networks.
\newblock {\em Proceedings of the National Academy of Sciences},
  113(36):9977--9982, 2016.

\bibitem{wasserman1994social}
Stanley Wasserman and Katherine Faust.
\newblock {\em Social network analysis: Methods and applications}, volume~8.
\newblock Cambridge university press, 1994.

\bibitem{newman2010networks}
Mark Newman.
\newblock {\em Networks: an introduction}.
\newblock Oxford university press, 2010.

\bibitem{holme2012temporal}
Petter Holme and Jari Saram{\"a}ki.
\newblock Temporal networks.
\newblock {\em Physics Reports}, 519(3):97--125, 2012.

\bibitem{tantipathananandh2007framework}
Chayant Tantipathananandh, Tanya Berger-Wolf, and David Kempe.
\newblock A framework for community identification in dynamic social networks.
\newblock In {\em Proceedings of the 13th ACM SIGKDD international conference
  on Knowledge discovery and data mining}, pages 717--726. ACM, 2007.

\bibitem{rosvall2014memory}
Martin Rosvall, Alcides~V Esquivel, Andrea Lancichinetti, Jevin~D West, and
  Renaud Lambiotte.
\newblock Memory in network flows and its effects on spreading dynamics and
  community detection.
\newblock {\em Nature Communications}, 5:4630, 2014.

\bibitem{xu2016representing}
Jian Xu, Thanuka~L Wickramarathne, and Nitesh~V Chawla.
\newblock Representing higher-order dependencies in networks.
\newblock {\em Science Advances}, 2(5):e1600028, 2016.

\bibitem{lerman2010centrality}
Kristina Lerman, Rumi Ghosh, and Jeon~Hyung Kang.
\newblock Centrality metric for dynamic networks.
\newblock In {\em Proceedings of the Eighth Workshop on Mining and Learning
  with Graphs}, pages 70--77. ACM, 2010.

\bibitem{liao2017ranking}
Hao Liao, Manuel~Sebastian Mariani, Matus Medo, Yi-Cheng Zhang, and Ming-Yang
  Zhou.
\newblock Ranking in evolving complex networks.
\newblock {\em Physics Reports}, 689:1--54, 2017.

\bibitem{li2017fundamental}
Aming Li, Sean~P Cornelius, Y-Y Liu, Long Wang, and A-L Barab{\'a}si.
\newblock The fundamental advantages of temporal networks.
\newblock {\em Science}, 358(6366):1042--1046, 2017.

\bibitem{read2008dynamic}
Jonathan~M Read, Ken~TD Eames, and W~John Edmunds.
\newblock Dynamic social networks and the implications for the spread of
  infectious disease.
\newblock {\em Journal of The Royal Society Interface}, 5(26):1001--1007, 2008.

\bibitem{scholtes2014causality}
Ingo Scholtes, Nicolas Wider, Ren{\'e} Pfitzner, Antonios Garas, Claudio~J
  Tessone, and Frank Schweitzer.
\newblock Causality-driven slow-down and speed-up of diffusion in non-markovian
  temporal networks.
\newblock {\em Nature Communications}, 5:5024, 2014.

\bibitem{holme2016temporal}
Petter Holme.
\newblock Temporal network structures controlling disease spreading.
\newblock {\em Physical Review E}, 94(2):022305, 2016.

\bibitem{fortunato2016community}
Santo Fortunato and Darko Hric.
\newblock Community detection in networks: A user guide.
\newblock {\em Physics Reports}, 659:1--44, 2016.

\bibitem{schaeffer2007graph}
Satu~Elisa Schaeffer.
\newblock Graph clustering.
\newblock {\em Computer Science Review}, 1(1):27--64, 2007.

\bibitem{mariani2019nestedness}
Manuel~Sebastian Mariani, Zhuo-Ming Ren, Jordi Bascompte, and Claudio~Juan
  Tessone.
\newblock Nestedness in complex networks: Observation, emergence, and
  implications.
\newblock {\em Physics Reports}, 813:1--90, 2019.

\bibitem{lynn2019physics}
Christopher~W Lynn and Danielle~S Bassett.
\newblock The physics of brain network structure, function and control.
\newblock {\em Nature Reviews Physics}, page~1, 2019.

\bibitem{holland1983stochastic}
Paul~W Holland, Kathryn~Blackmond Laskey, and Samuel Leinhardt.
\newblock Stochastic blockmodels: First steps.
\newblock {\em Social networks}, 5(2):109--137, 1983.

\bibitem{mahmood2016using}
Arif Mahmood, Michael Small, Somaya~Ali Al-Maadeed, and Nasir Rajpoot.
\newblock Using geodesic space density gradients for network community
  detection.
\newblock {\em IEEE Transactions on Knowledge and Data Engineering},
  29(4):921--935, 2016.

\bibitem{newman2004finding}
Mark~EJ Newman and Michelle Girvan.
\newblock Finding and evaluating community structure in networks.
\newblock {\em Physical Review E}, 69(2):026113, 2004.

\bibitem{dorogovtsev2000evolution}
Sergey~N Dorogovtsev and Jos{\'e} Fernando~F Mendes.
\newblock Evolution of networks with aging of sites.
\newblock {\em Physical Review E}, 62(2):1842, 2000.

\bibitem{medo2011temporal}
Mat{\'u}{\v{s}} Medo, Giulio Cimini, and Stanislao Gualdi.
\newblock Temporal effects in the growth of networks.
\newblock {\em Physical Review Letters}, 107(23):238701, 2011.

\bibitem{golosovsky2017growing}
Michael Golosovsky and Sorin Solomon.
\newblock Growing complex network of citations of scientific papers: Modeling
  and measurements.
\newblock {\em Physical Review E}, 95(1):012324, 2017.

\bibitem{leicht2007large}
Elizabeth~A Leicht, Gavin Clarkson, Kerby Shedden, and Mark~EJ Newman.
\newblock Large-scale structure of time evolving citation networks.
\newblock {\em The European Physical Journal B}, 59(1):75--83, 2007.

\bibitem{chen2010community}
P~Chen and Sidney Redner.
\newblock Community structure of the physical review citation network.
\newblock {\em Journal of Informetrics}, 4(3):278--290, 2010.

\bibitem{stella2018bots}
Massimo Stella, Emilio Ferrara, and Manlio De~Domenico.
\newblock Bots increase exposure to negative and inflammatory content in online
  social systems.
\newblock {\em Proceedings of the National Academy of Sciences},
  115(49):12435--12440, 2018.

\bibitem{mucha2010community}
Peter~J Mucha, Thomas Richardson, Kevin Macon, Mason~A Porter, and Jukka-Pekka
  Onnela.
\newblock Community structure in time-dependent, multiscale, and multiplex
  networks.
\newblock {\em Science}, 328(5980):876--878, 2010.

\bibitem{granell2015benchmark}
Clara Granell, Richard~K Darst, Alex Arenas, Santo Fortunato, and Sergio
  G{\'o}mez.
\newblock Benchmark model to assess community structure in evolving networks.
\newblock {\em Physical Review E}, 92(1):012805, 2015.

\bibitem{bazzi2016community}
Marya Bazzi, Mason~A Porter, Stacy Williams, Mark McDonald, Daniel~J Fenn, and
  Sam~D Howison.
\newblock Community detection in temporal multilayer networks, with an
  application to correlation networks.
\newblock {\em Multiscale Modeling \& Simulation}, 14(1):1--41, 2016.

\bibitem{chai2016functional}
Lucy~R Chai, Marcelo~G Mattar, Idan~Asher Blank, Evelina Fedorenko, and
  Danielle~S Bassett.
\newblock Functional network dynamics of the language system.
\newblock {\em Cerebral Cortex}, 26(11):4148--4159, 2016.

\bibitem{krings2012effects}
Gautier Krings, M{\'a}rton Karsai, Sebastian Bernhardsson, Vincent~D Blondel,
  and Jari Saram{\"a}ki.
\newblock Effects of time window size and placement on the structure of an
  aggregated communication network.
\newblock {\em EPJ Data Science}, 1(1):4, 2012.

\bibitem{darst2016detection}
Richard~K Darst, Clara Granell, Alex Arenas, Sergio G{\'o}mez, Jari
  Saram{\"a}ki, and Santo Fortunato.
\newblock Detection of timescales in evolving complex systems.
\newblock {\em Scientific Reports}, 6:39713, 2016.

\bibitem{girvan2002community}
Michelle Girvan and Mark~EJ Newman.
\newblock Community structure in social and biological networks.
\newblock {\em Proceedings of the National Academy of Sciences},
  99(12):7821--7826, 2002.

\bibitem{arenas2007size}
Alex Arenas, Jordi Duch, Alberto Fern{\'a}ndez, and Sergio G{\'o}mez.
\newblock Size reduction of complex networks preserving modularity.
\newblock {\em New Journal of Physics}, 9(6):176, 2007.

\bibitem{malliaros2013clustering}
Fragkiskos~D Malliaros and Michalis Vazirgiannis.
\newblock Clustering and community detection in directed networks: A survey.
\newblock {\em Physics Reports}, 533(4):95--142, 2013.

\bibitem{blondel2008fast}
Vincent~D Blondel, Jean-Loup Guillaume, Renaud Lambiotte, and Etienne Lefebvre.
\newblock Fast unfolding of communities in large networks.
\newblock {\em Journal of Statistical Mechanics: Theory and Experiment},
  2008(10):P10008, 2008.

\bibitem{ren2018randomizing}
Zhuo-Ming Ren, Manuel~Sebastian Mariani, Yi-Cheng Zhang, and Mat{\'u}{\v{s}}
  Medo.
\newblock Randomizing growing networks with a time-respecting null model.
\newblock {\em Physical Review E}, 97(5):052311, 2018.

\bibitem{lancichinetti2008benchmark}
Andrea Lancichinetti, Santo Fortunato, and Filippo Radicchi.
\newblock Benchmark graphs for testing community detection algorithms.
\newblock {\em Physical Review E}, 78(4):046110, 2008.

\bibitem{peel2017ground}
Leto Peel, Daniel~B Larremore, and Aaron Clauset.
\newblock The ground truth about metadata and community detection in networks.
\newblock {\em Science Advances}, 3(5):e1602548, 2017.

\bibitem{spitz2015breaking}
Andreas Spitz and Michael Gertz.
\newblock Breaking the news: Extracting the sparse citation network backbone of
  online news articles.
\newblock In {\em Proceedings of the 2015 IEEE/ACM International Conference on
  Advances in Social Networks Analysis and Mining}, pages 274--279. ACM, 2015.

\bibitem{parolo2015attention}
Pietro Della~Briotta Parolo, Raj~Kumar Pan, Rumi Ghosh, Bernardo~A Huberman,
  Kimmo Kaski, and Santo Fortunato.
\newblock Attention decay in science.
\newblock {\em Journal of Informetrics}, 9(4):734--745, 2015.

\bibitem{newman2016equivalence}
MEJ Newman.
\newblock Equivalence between modularity optimization and maximum likelihood
  methods for community detection.
\newblock {\em Physical Review E}, 94(5):052315, 2016.

\bibitem{pamfil2018relating}
A~Roxana Pamfil, Sam~D Howison, Renaud Lambiotte, and Mason~A Porter.
\newblock Relating modularity maximization and stochastic block models in
  multilayer networks.
\newblock {\em arXiv preprint arXiv:1804.01964}, 2018.

\bibitem{ghasemian2016detectability}
Amir Ghasemian, Pan Zhang, Aaron Clauset, Cristopher Moore, and Leto Peel.
\newblock Detectability thresholds and optimal algorithms for community
  structure in dynamic networks.
\newblock {\em Physical Review X}, 6(3):031005, 2016.

\bibitem{lancichinetti2012consensus}
Andrea Lancichinetti and Santo Fortunato.
\newblock Consensus clustering in complex networks.
\newblock {\em Scientific Reports}, 2:336, 2012.

\bibitem{medo2014statistical}
Mat{\'u}{\v{s}} Medo.
\newblock Statistical validation of high-dimensional models of growing
  networks.
\newblock {\em Physical Review E}, 89(3):032801, 2014.

\bibitem{holme2015modern}
Petter Holme.
\newblock Modern temporal network theory: a colloquium.
\newblock {\em The European Physical Journal B}, 88(9):234, 2015.

\bibitem{easley2010networks}
David Easley and Jon Kleinberg.
\newblock {\em Networks, crowds, and markets: Reasoning about a highly
  connected world}.
\newblock Cambridge University Press, 2010.

\bibitem{zhou2010solving}
Tao Zhou, Zolt{\'a}n Kuscsik, Jian-Guo Liu, Mat{\'u}{\v{s}} Medo,
  Joseph~Rushton Wakeling, and Yi-Cheng Zhang.
\newblock Solving the apparent diversity-accuracy dilemma of recommender
  systems.
\newblock {\em Proceedings of the National Academy of Sciences},
  107(10):4511--4515, 2010.

\bibitem{fortunato2010community}
Santo Fortunato.
\newblock Community detection in graphs.
\newblock {\em Physics Reports}, 486(3):75--174, 2010.
\end{thebibliography}
\end{document}